\documentclass[11pt]{article}
\usepackage[T1]{fontenc}
\usepackage{array}
\usepackage{amsmath}
\usepackage{amssymb}
\usepackage{graphicx}
\usepackage{booktabs}
\usepackage[authoryear]{natbib}
\usepackage[unicode=true,
 bookmarks=false,
 breaklinks=false,pdfborder={0 0 1},backref=section,colorlinks=false]
 {hyperref}

\makeatletter

\providecommand{\tabularnewline}{\\}

\@ifundefined{date}{}{\date{}}
\usepackage{latexsym}
\usepackage{multirow}\usepackage{bm}%\usepackage[latin9]{inputenc}
\usepackage{url}% not crucial - just used below for the URL 

\usepackage{enumerate}

\usepackage{colortbl}
\usepackage{subcaption}

\usepackage{dcolumn}
\newcolumntype{.}{D{.}{.}{-1}}
\newcolumntype{d}[1]{D{.}{.}{#1}}
\usepackage{theorem}
\newtheorem{assumption}{Assumption}\newtheorem{theorem}{Theorem}\newtheorem{remark}{Remark}\newtheorem{model}{Model}\newtheorem{proof}{Proof}
\usepackage{rotating}

\usepackage{arydshln}

\usepackage[compact]{titlesec}

\allowdisplaybreaks

\newcommand{\spacingset}[1]{\renewcommand{\baselinestretch}%
{#1}\small\normalsize}

\newcommand{\pr}{P}

\newcommand{\E}{\mathbb{E}}

\makeatother

\begin{document}
\title{\textbf{Doubly robust omnibus sensitivity analysis of externally controlled trials with intercurrent events}}
\author{Chenyin Gao$^{1}$, Xiang Zhang$^{2}$ Shu Yang$^{1}$\thanks{Corresponding author: syang24@ncsu.edu}\\
$^{1}$Department of Statistics, North Carolina State University\\
$^{2}$Medical Affairs and HTA Statistics, CSL Behring
}
\spacingset{1.5} 
\maketitle

\begin{abstract}
Externally controlled trials are crucial in clinical development when randomized controlled trials are unethical or impractical. These trials consist of a full treatment arm with the experimental treatment and a full external control arm. However, they present significant challenges in learning the treatment effect due to the lack of randomization and a parallel control group. Besides baseline incomparability, outcome mean non-exchangeability, caused by differences in conditional outcome distributions between external controls and counterfactual concurrent controls, is infeasible to test and may introduce biases in evaluating the treatment effect. Sensitivity analysis of outcome mean non-exchangeability is thus critically important to assess the robustness of the study's conclusions against such assumption violations. Moreover, intercurrent events, which are ubiquitous and inevitable in clinical studies, can further confound the treatment effect and hinder the interpretation of the estimated treatment effects. This paper establishes a semi-parametric framework for externally controlled trials with intercurrent events, offering doubly robust and locally optimal estimators for primary and sensitivity analyses. We develop an omnibus sensitivity analysis that accounts for both outcome mean non-exchangeability and the impacts of intercurrent events simultaneously, ensuring root-n consistency and asymptotic normality under specified conditions. The performance of the proposed sensitivity analysis is evaluated in simulation studies and a real-data problem.
\end{abstract}
%  As usual, the \maketitle command creates the title and author/affiliations
%  display 

\maketitle
\noindent%
\vfill

\newpage

\section{Introduction\label{sec:intro} }

\subsection{Why consider external controls}

In medical research, the gold standard for evaluating new treatments has been
\textit{randomized controlled trials} (RCTs). Regulatory bodies often
require these rigorously controlled clinical studies to validate
the effectiveness and safety of these treatments for specific patient
groups. Despite the high regard for randomized, double-blind trials,
they may not always be viable or ethical, particularly in rare or
severe diseases with limited treatment options. Recruiting enough
participants for such trials is often difficult, and using placebos in situations when alternative effective treatments are available may be unethical or impractical.

In these cases, \textit{single-arm trials} (SATs) can be a practical
alternative, though they come with limitations. Without
direct comparison data for untreated subjects, researchers are compelled to infer these outcomes from external sources like previous studies, or real-world databases. These are referred to
as \textit{externally controlled trials} (ECTs). However, data from past trials might not always be relevant due to changes in the treatment
landscape and patient demographics, reducing its evidence capacity
compared with RCTs. Recently, real-world data has gained popularity for external control arms due to its accessibility and contemporaneity with treatment groups. 

\subsection{Estimand considerations with external controls }

The Food and Drug Administration's (FDA) latest draft guidelines for
natural history studies in rare disease drug development emphasized five critical concerns with using external controls \citep{FDA2023}. These concerns, outlined in Table
\ref{tab:Key-considerations}, may introduce
biases in research findings when real-world data is utilized as a
control arm in ECTs. For a more structured analysis of these biases,
we classified them into two principal categories: baseline incomparability
and outcome mean non-exchangeability. Each category is characterized by unique
mechanisms for introducing bias, as detailed in Table \ref{tab:Key-considerations}.

\begin{table}
\begin{centering}
\caption{\label{tab:Key-considerations}Key considerations of using external
controls}
\begin{tabular}{p{.25\textwidth}p{.7\textwidth}}
\toprule
\multicolumn{2}{l}{\textbf{(Observed) baseline incomparability}}  \\ 
\midrule
Covariate distribution shift & Systematic differences exist in the baseline characteristics of patients in external studies compared to those in SATs.                                \\
\midrule
\multicolumn{2}{l}{\textbf{(Unobserved) outcome mean non-exchangeability}} \\
Unmeasured confounding       & External control data might not capture the same detailed patient information as SATs, leading to biases from unknown or unmeasured factors.           \\
Lack of concurrency/ temporal bias         & External control data and SATs may be collected in different time periods or under varying healthcare settings.                                        \\
Measurement error            & There may exist potential inconsistencies in how patient information is collected and recorded, termed here as measurement errors in covariates.       \\
Outcome validity             & Methods of measuring outcomes in external data sources might differ from those used in SATs, or the outcomes might not be clearly defined or reliable. \\
\midrule
\multicolumn{2}{l}{\textbf{Proper estimands}}\\
Intercurrent events          & Intercurrent events, such as stopping medication and/or adding rescue therapy, can confound the causal effects of the randomized treatment.            \\ \hline
\end{tabular}
\end{centering}
\end{table}

Besides the challenges of non-exchangeability of the external controls in clinical trials,
managing intercurrent events like participant dropout, non-compliance or premature termination of therapy adds more complexities.
Typically, the Missingness at Random (MAR) framework is adopted after
the absence of outcomes following these events.
While the MAR assumption is often deemed plausible, it remains unverifiable and may lack practical applicability since it assumes all participants
persist with the study medication, addressing only a theoretical
aspect of the treatment effect, as noted in the guidelines \citep{international2019addendum}. A more plausible assumption would be that the treatment effect may quickly fade away, leading to the Missing Not at Random (MNAR) pattern for the intercurrent events. The MNAR assumption can also be put forward for the sensitivity analysis, suggesting that any treatment effect observed while a participant was active in the study is nullified upon their discontinuation. To evaluate this effect, the
implementation of Control-based Imputation (CBI) models has been proposed, offering a nuanced and realistic assessment of treatment outcomes in clinical trials.

\subsection{Primary analysis and doubly robust omnibus sensitivity analysis}\label{sec:contribution}

Numerous statistical methodologies have been developed to address
the potential biases from using external controls in
ECTs. Most of these methods utilize techniques like propensity
score stratification, matching, or weighting to mitigate selection
bias by balancing baseline covariates between external controls and SATs. Nonetheless, the outcome exchangeability of external controls cannot be verified with observed data due to
the absence of concurrent controls. Therefore, sensitivity
analyses become essential to evaluate the impact of potential violations of outcome
non-exchangeability and to understand the effects of intercurrent events under different realistic scenarios.

In this context, we introduce a sensitivity analysis framework tailored for ECTs with intercurrent events. This framework encompasses models like the tilting models, which control the degree of outcome mean non-exchangeability in external controls and alterations in outcome distributions after intercurrent events. Central to our framework is the establishment of identification results, the development of efficient influence functions (EIFs), and EIF-motivated tilting estimators under sensitivity models, which jointly capture the effect changes due to outcome mean non-exchangeability and intercurrent events. These EIF-motivated estimators have several advantageous statistical properties, such as local efficiency, double robustness, and asymptotic normality. By analytically establishing the conditions for desirable asymptotic properties, our estimator allows flexible models for nuisance parameters while maintaining root-n consistency (Theorem \ref{thm:EIF-tilting-1}). Therefore, our major contribution is to derive the locally efficient estimator for evaluating the treatment effect under the sensitivity models and to jointly assess the robustness and reliability of multiple assumptions in ECTs with intercurrent events in a more efficient and flexible manner.

Our paper is organized as follows: Section \ref{sec:related_work} presents a brief overview of sensitivity analysis. In Section \ref{sec:1time} introduces the notation and develops the semi-parametric efficient estimator for the primary analysis. Section \ref{sec:Sensitivity-analysis} details the tilting sensitivity models and the efficient inference for the sensitivity analysis. Another efficient estimator for the CBI sensitivity analysis under the Jump-to-Reference (J2R) model is presented in the Supplementary Materials. Section \ref{sec:calibrating}, discusses one practical method for choosing the sensitivity parameters by bounding their impacts. Extensive simulation studies for both primary and sensitivity analyses are presented in Section \ref{sec:simu}. Section \ref{sec:real} illustrates our approach with an antidepressant trial, and Section \ref{sec:discuss} concludes with a discussion.

\section{Related Works}\label{sec:related_work}

Before we delve into the proposed sensitivity analysis framework,
we provide a review of sensitivity analysis methods. Causal inference
involving observational studies usually requires no unmeasured confounding
assumption, that is the treatment assignment is ignorable conditional
on a set of covariates \citep{rosenbaum1983central,imbens2015causal}.
Yet, claiming the absence of confounders in the treatment-outcome relationship is untestable and often implausible in practice. Thus,
it is advised to conduct a series of sensitivity analyses assessing
how robust the causal findings are against the plausible violations
of the unconfoundedness \citep{faries2023real}. The problem of sensitivity analyses has been
studied in a variety of fields with the earliest work in \citet{cornfield1959smoking},
which is later extended in \citet{rosenbaum1983assessing,rosenbaum1987sensitivity,imbens2003sensitivity}. However, one concerning issue regarding this
framework is that it demands a specific parametric assumption on the unmeasured confounder $U$. In many cases, sensitivity
analyses can be quite fragile against the model misspecification of $U$ as shown in \citet{zhang2019semiparametric}.  

% In particular, the following general formulation is considered: 
% \begin{align}
%  & U\mid X\sim f_{U}(u\mid X),\label{eq:U_cont}\\
%  & \E(A\mid X,U)=g_{1}^{-1}(\alpha^{\intercal}X+\zeta^{a}U),\label{eq:A_cont}\\
%  & \E(Y\mid A,X,U)=g_{2}^{-1}(\tau A+\beta^{\intercal}X+\zeta^{y}U),\label{eq:Y_cont}
% \end{align}
% where $U$ is the unmeasured confounder, $X$ is a vector of measured
% covariates including intercept, $A$ is the binary treatment, $Y$
% is the outcome of interest, and $(\zeta^{a},\zeta^{y})$ are two sensitivity
% parameters that govern the impact of $U$ on $A$ and $Y$. It is
% assumed that $(A,Y)$ belongs to exponential family with the canonical
% link functions $(g_{1},g_{2})$, and $f_{U}(u\mid X)$ is a user-specified
% parametric distribution. 

To circumvent modeling the conditional (or marginal) distribution
for the unmeasured confounder, a plethora of sensitivity approaches
have been proposed, which preserves the critical elements in \citet{rosenbaum1983assessing,imbens2003sensitivity}
and allows for flexible strategies to model the distribution of $U$.
\citet{zhang2019semiparametric} leverages the modern semi-parametric
theory to obtain a consistent estimation in a model while placing
no distributional assumptions on $U$. The idea of partial misspecification
of the nuisance parameters endows the framework with an unrestricted
law of latent variables and has been similarly considered in the context
of measurement error \citep{tsiatis2004locally}, mixed models \citep{garcia2016optimal},
and statistical genetics \citep{allen2005locally}. Another line of
work to tackle this problem is based on the idea of ``omitted-variable''
bias (OVB), which can be computed easily without needs to specify
the parametric form of the potential unobserved confounding. The idea
of general bias formula is introduced in \citet{vanderweele2011bias},
and is further extended in \citet{cinelli2020making}
with more flexibility and robustness.

As illustrated so far, most of the work blurs the line between sensitivity analysis and model checking by introducing sensitivity parameters under stringent parametric assumptions. For example, \cite{little2019statistical} suggest that the ignorability assumption in \cite{heckman1979sample} can be tested as the result of their Gaussian parametric assumption, thereby inducing testable implications of the untestable ignorability assumption. Moreover,
the lack of such ``clean'' separation requires that the observed model
to be refit for each setting of the sensitivity parameters, which
will be an onerous task as modern non-parametric models may be adopted
to fit the potential outcomes. To address this concern, \citet{robins2000sensitivity}
propose and extended by \citet{franks2019flexible,nabi2024semiparametric}
to use the ``tilting'', or ``selection'' function to decouple
the sensitivity analyses from the observed data model. Typically,
such sensitivity analysis specification does not impose any parametric assumptions on the distribution of the observed data or the unmeasured confounder, but only on a relaxed version of the unconfoundedness
assumption:
\begin{align}
f\{Y(a),A=1-a\mid X\} & =f\{Y(a),A=a\mid X\}\frac{f\{A=1-a\mid Y(a),X;\psi\}}{f\{A=a\mid Y(a),X;\psi\}},\label{eq:tilting}
\end{align}
where $Y(a)$ is the potential outcome under treatment $a$, $A$ is the treatment assignment, and $X$ is the baseline covariates. Here, the first term constitutes the observed data density, while the
second term is the selection function governed by the sensitivity
parameter $\psi$. Other approaches in \cite{blackwell2014selection,yang2018sensitivity}
take a different track by representing the confounding as a function
of the observed covariates, and describe the conditional potential
outcomes difference varied by the treatment assignment as $q(a,X;\alpha)=\E\{Y(a)\mid A=a,X\}-\E\{Y(a)\mid A=1-a,X\}$, where the confounding function $q$ is characterized by the single
sensitivity parameter $\alpha$. As the observed data distribution
is free of the sensitivity parameters ($\psi$ or $\alpha$), it achieves
the ``clean'' factorization of the identified and unidentified parts
of the sensitivity analysis framework. \citet{veitch2020sense}, extending
from \citet{imbens2003sensitivity}, posits a probabilistic model
to bypass the need to specify any distributional assumption on $U$,
which also decouples the sensitivity analyses from the observed data
and leads to tractable bias calculations.

\section{Single-arm trial data with external controls: primary analysis \label{sec:1time}}

To ground ideas, we first focus on cross-sectional studies. Let $S=1$
denote trial participation and the trial data consist of $\{V_{i}=(X_{i},A_{i},R_{i},Y_{i},S_{i}=1):i\in\mathcal{R}\}$, where $R=1$ indicates the absence of intercurrent events (e.g., treatment discontinuation) during follow-up and zero otherwise. Let $S=0$ denote the external participation and the external
data consist of $\{V_{i}=(X_{i},A_{i},R_{i},Y_{i}, S_{i}=0):i\in\mathcal{E}\}$.
Assume $(X_{i},A_{i},R_{i},Y_{i},S_{i})$ are independent and identically
distributed, and we omit the subscript $i$ for the simplicity of
notation. Let $V=(X,A,R,Y,S)$ be the random vector of all observed
variables and follow the observed data distribution $P_{0}$. The
\textit{treatment policy} strategy designates a treatment effect estimand
that measures the total effect of the treatment assignment and the
intercurrent event on the outcome. Therefore, to define the estimand
unambiguously, we extend the causal framework in \citet{lipkovich2020causal}
and introduce the potential outcomes framework for $R$ and $Y$. Denote $R(a)$ as the potential indicator for the absence of intercurrent events under treatment $a$, $Y(a, r)$ as the potential outcome under treatment $a$ with intercurrent event status $r$, and $Y(a) = Y\{a, R(a)\}$. 

The estimand of interest is the average treatment effect (ATE) for
the trial, defined as 
\[
\tau=\E[Y\{1,R(1)\}-Y\{0,R(0)\}\mid S=1] = \E\{Y(1)-Y(0)\mid S=1\},
\]
which is a \textit{treatment policy} estimand, as the occurrence of intercurrent events is considered irrelevant. However, due to the missing outcomes following these events, the \textit{treatment policy} strategy cannot be implemented for intercurrent events that are terminal, such as treatment discontinuation. Table \ref{tab:Key-assumptions}(A) outlines several key causal assumptions
to identify the treatment effect for the primary analysis.

\begin{table}
\centering
\scriptsize
\caption{\label{tab:Key-assumptions} Lists of (A) key assumptions for primary analysis (B) necessary notation}
\begin{tabular}{p{.35\textwidth}p{.6\textwidth}}
\hline 
(A) \textbf{Assumptions} & \textbf{Details}\tabularnewline
\hline 
1. Causal consistency &  $R=R(A),$ and $Y=Y\left\{ A,R(A)\right\}$. \\
2. Positivity & $P(S=1\mid X) > 0$ and $P(R=1\mid X, S=s) > 0$ for $s=0, 1$.\\
3. Outcome exchangeability of the external controls & $\E\{Y(0)\mid X,S=1\}=\E\{Y(0)\mid X,S=0\}=\mu_{0}(X)$.\tabularnewline
4. Ignorability of intercurrent event for SAT data & $R(a)\perp Y(a,r)\mid X,S=1$, for all $a,r$.\tabularnewline
5. Ignorability of intercurrent event for external controls & $R(a)\perp Y(a,r)\mid X,S=0$, for all $a,r$.\tabularnewline
\hline 
\hline 
(B) \textbf{Formula} & \textbf{Details}\tabularnewline
\hline 
$\pi_{S}(X)$ & participation propensity, defined as $\pi_{S}(X)=P(S=1\mid X)$\tabularnewline
$q_{S}(X)$ & participation propensity density ratio, defined as $q_{S}(X)=\pi_{S}(X)/\{1-\pi_{S}(X)\}$\tabularnewline
$\pi_{R_{s}}(X)$ & propensity of  not having intercurrent event for $s=0,1$, defined as $\pi_{R_{s}}(X)=P(R=1\mid X,S=s)$\tabularnewline
$q_{R_{s}}(X)$ & propensity density ratio of intercurrent event for $s=0,1$, defined
as $q_{R_{s}}(X)=\{1-\pi_{R_{1}}(X)\}/\pi_{R_{1}}(X)$\tabularnewline
$\mu_{s}(X)$ & outcome means for $s=0,1$, defined as $\mu_{s}(X)=\E(Y\mid X,S=s,R=1)$\tabularnewline
\hline 
$c(X;\gamma_{R_{0}})$, $c(X;\gamma_{R_{1}})$,

$c(X;\gamma_{S})$, $c(X;\gamma_{S}+\gamma_{R_{0}})$ & 
Normalizing terms, defined as $c(X;\gamma_{R_{0}})=\E[\exp\{\gamma_{R_{0}},Y(0)\}\mid X,S=0,R=1]$, $c(X;\gamma_{R_{1}})=\E[\exp\{\gamma_{R_{1}}Y(1)\}\mid X,S=1,R=1)]$, $c(X;\gamma_{S})=\E[\exp\{\gamma_{S}Y(0)\}\mid X,S=0,R=1]$, $c(X;\gamma_{S}+\gamma_{R_{0}})=\E[\exp\{(\gamma_{S}+\gamma_{R_{0}})Y(0)\}\mid X,S=0,R=1]$\tabularnewline
\hline 
$b(X;\gamma_{R_{0}})$, $b(X;\gamma_{R_{1}})$,

$b(X;\gamma_{S})$, $b(X;\gamma_{S}+\gamma_{R_{0}})$ & Tilted outcome means, defined as $b(X;\gamma_{R_{0}})=\E[Y\exp\{\gamma_{R_{0}},Y(0)\}\mid X,S=0,R=1]$,
$b(X;\gamma_{R_{1}})=\E[Y\exp\{\gamma_{R_{1}}Y(1)\}\mid X,S=1,R=1]$,
$b(X;\gamma_{S})=\E[Y\exp\{\gamma_{S}Y(0)\}\mid X,S=0,R=1]$,

$b(X;\gamma_{S}+\gamma_{R_{0}})=\E[Y\exp\{(\gamma_{S}+\gamma_{R_{0}})Y(0)\}\mid X,S=0,R=1]$\tabularnewline
\hline 
$d(X;\gamma_{R_{0}},\gamma_{S})$, $e(X;\gamma_{R_{0}},\gamma_{S})$ & $d(X;\gamma_{R_{0}},\gamma_{S})=\pi_{R_{0}}(X)b(X;\gamma_{S})c(X;\gamma_{R_{0}})+\{1-\pi_{R_{0}}(X)\}b(X;\gamma_{S}+\gamma_{R_{0}})$,

$e(X;\gamma_{R_{0}},\gamma_{S})=\pi_{R_{0}}(X)c(X;\gamma_{S})c(X;\gamma_{R_{0}})+\{1-\pi_{R_{0}}(X)\}c(X;\gamma_{S}+\gamma_{R_{0}})$\tabularnewline
\hline 
\end{tabular}
\end{table}
Assumptions 1 and 2 are standard causal assumptions for identification \citep{rosenbaum1983central}. Assumption 1 maps the potential outcomes to the observed data, and Assumption 2 ensures that each participant has a positive probability of being recruited into the SAT or external controls, and not having intercurrent events. Assumption 3 states that the conditional expectation of $Y(0)$ is
the same for the trial and the external controls. Assumptions
4 and 5 imply that the intercurrent events occur at random for SAT
and the external controls, respectively. These assumptions are satisfied
if the covariates $X$ capture all the confounding variables. Take a weight-loss trial for an example, where the intercurrent events are the non-compliance with the prescribed diet. If we assume that the covariates, such as age, baseline weight, and other lifestyle factors, capture all the confounding variables, it follows that given the same covariates for two participants, they have the same likelihood of being non-compliant with the diet, regardless of their weight loss. Therefore, we can conclude that the weight loss will not be affected by the occurrence of the non-compliance conditional on these covariates, and the post-intercurrent event outcomes are exchangeable to the observed outcomes.

Theorem \ref{Thm:id_primary} provides three identification formulas for $\tau$ under the assumptions in Table
\ref{tab:Key-assumptions}(A), and Table \ref{tab:Key-assumptions}(B)
summarizes the necessary models for the identification.

\begin{theorem}[Identification]\label{Thm:id_primary} Under the assumptions in Table
\ref{tab:Key-assumptions}, $\tau$ is identifiable by
\begin{enumerate}
\item [(a)] trial participation propensity and outcome means:  
$$
\tau=\frac{\E\{\pi_{S}(X)\mu_{1}(X)-\pi_{S}(X)\mu_{0}(X)\}}{P(S=1)}.
$$
\item [(b)] trial participation propensity and response propensity:
\[
\tau=\frac{1}{P(S=1)}\E\left\{ \frac{SRY}{\pi_{R_{1}}(X)}-\frac{(1-S)Rq_{S}(X)Y}{\pi_{R_{0}}(X)}\right\} .
\]
\item [(c)] response propensity and outcome mean:
\begin{align*}
\tau & =\frac{1}{P(S=1)}\E\left\{ S\pi_{R_{1}}(X)Y+S\{1-\pi_{R_{1}}(X)\}\mu_{1}(X)-S\mu_{0}(X)\right\} .
\end{align*}
\end{enumerate}
\end{theorem}
We give some intuitions behind these identification formulas. Theorem
\ref{Thm:id_primary}(a) describes that the individual treatment
effect given the covariates $X$ is $\mu_{1}(X)-\mu_{0}(X)$. Taking
the expectation over the trial population yields the identification
for ATE. Theorem \ref{Thm:id_primary}(b) can be understood as a
transportability problem. The first term, corresponding to $\pi_{S}(X)\mu_{1}(X)$,
adjusts the outcomes from SAT $SY$ by $R/\pi_{R_{1}}(X)$, which
weights the observed subjects by their response propensity. The second
term, corresponding to $\pi_{S}(X)\mu_{0}(X)$, adjusts the outcomes
$(1-S)Y$ from the external controls by $Rq_{S}(X)/\pi_{R_{0}}(X)$,
which transport the external controls to the trial via the density
ratio $q_{S}(X)$ after response propensity weighting. In Theorem
\ref{Thm:id_primary}(c), it predicts the outcomes for the treatment
group by $\pi_{R_{1}}(X)Y+\{1-\pi_{R_{1}}(X)\}\mu_{1}(X)$, which
imputes the post-intercurrent event outcomes by $\mu_{1}(X)$. Similarly,
it predicts the outcomes of the external controls by $\mu_{0}(X)$.
The difference between these two predictions marginalized over the
trial population quantifies the ATE.

Based on these identification formulas, infinitely many estimators can be constructed. To develop a more principled estimator, we derive the EIF for $\tau$ and the resulting EIF-motivated tilting estimator achieves the rate double robustness, local efficiency, and asymptotic normality. The details of the estimator and these properties are relegated to Theorems \ref{Thm:eif0} and \ref{Thm:eif0-1} in the Supplementary Materials.

\section{Sensitivity analysis under tilting models \label{sec:Sensitivity-analysis}}

\subsection{Assumptions and a graphical representation}

Assumptions 3 to 5 in Table \ref{tab:Key-assumptions} are critical for
the identification of $\tau$. However,
these assumptions may be subject to violations in practice and are
unverifiable based on the observed data. Here, we develop the tilting
sensitivity models to assess whether the primary analysis result is
sensitive to the violation of these assumptions. 

\begin{model}[Tilting sensitivity models]\label{assump:tilting-MNAR}
Assume that the tilting models for EC outcome mean non-exchangeability and
the effects of intercurrent events are 
\begin{align*}
 & {\rm d}F\{Y(s,0)\mid X,S=s,R=0\}={\rm d}F\{Y(s,1)\mid X,S=s,R=1\}\frac{\exp\{\gamma_{R_{s}}Y(s)\}}{c(X;\gamma_{R_{s}})},\\
 % & {\rm d}F\{Y(1,0)\mid X,S=1,R=0\}={\rm d}F\{Y(1,1)\mid X,S=1,R=1\}\frac{\exp\{\gamma_{R_{1}}Y(1)\}}{c(X;\gamma_{R_{1}})},\\
 & {\rm d}F\{Y(0)\mid X,S=1\}={\rm d}F\{Y(0)\mid X,S=0\}\frac{\exp\{\gamma_{S}Y(0)\}}{\E[\exp\{\gamma_{S}Y(0)\}\mid X,S=0]},
\end{align*}
for $s=0,1$, where the normalizing terms $c(X;\gamma_{R_s})$ are defined in Table \ref{tab:Key-assumptions}(B).
\end{model}

\begin{figure}[htbp]
\centering
\includegraphics[width=.8\linewidth]{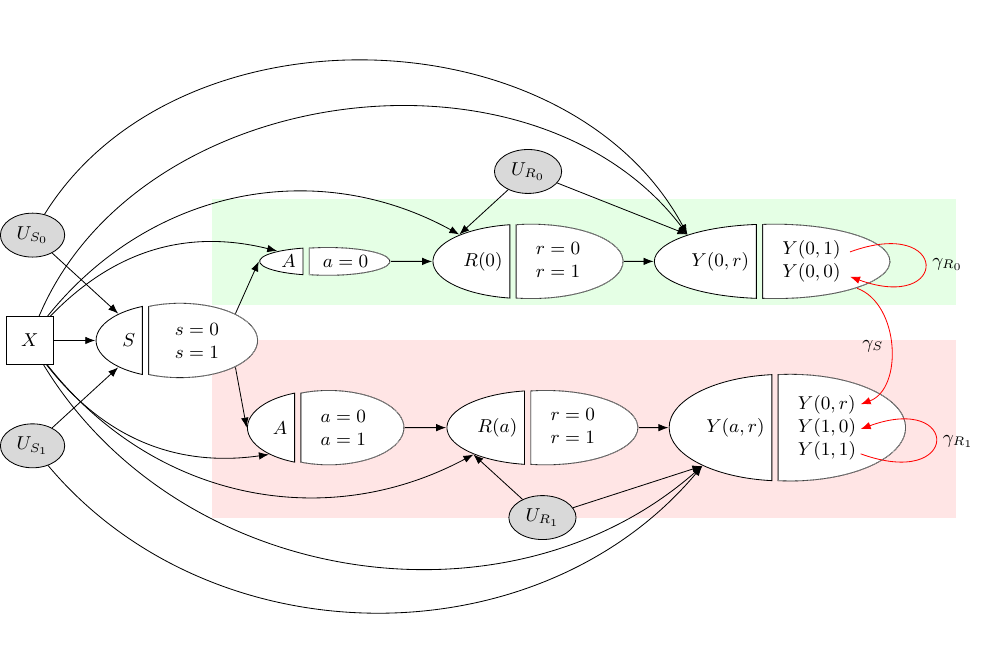}
\caption{\label{fig:plot-DAG-tilting} Schematic plot of the tilting sensitivity models subject to unmeasured confounders
$U_{S}$, $U_{R_{0}}$ and $U_{R_{1}}$.}
\end{figure}

Model \ref{assump:tilting-MNAR} posits that each unobserved
outcome distribution is a ``tilted version'' of the observed outcomes,
where $\gamma_{S}$, $\gamma_{R_{0}}$, and $\gamma_{R_{1}}$ are
treated as the sensitivity parameters, entailing the level of EC outcome
non-exchangeability and the effect of intercurrent events within each arm; see Figure \ref{fig:plot-DAG-tilting} for an illustration when Assumptions 3 to 5 in Table \ref{tab:Key-assumptions} are violated due to unmeasured confounders $U_{S}$, $U_{R_{0}}$ and $U_{R_{1}}$ under the tilting sensitivity models. 

Given negative (or positive) sensitivity parameters, the unobserved outcome distribution is tilted to the left (or right) relative to the distribution of observed outcomes, with smaller (or larger) values receiving greater weight. For example, if $\gamma_S$ is smaller (or larger) than zero, Model \ref{assump:tilting-MNAR} implies that trial participants, if untreated, tend to have smaller (or larger) outcomes compared to the external controls, given the same covariates. Similarly, if $\gamma_{R_0}$ and $\gamma_{R_1}$ are both smaller (or larger) than zero, Model \ref{assump:tilting-MNAR} implies that participants with intercurrent events tend to have smaller (or larger) outcomes compared to those without such events, given the same covariates. In particular, $\gamma_{S}=0$ leads to Assumption 3, $\gamma_{R_{0}}=0$
leads to Assumption 4, and $\gamma_{R_{1}}=1$ leads to Assumption
5.

\begin{remark}[Logistic selection]\label{rmk:logic_selection}

The tilting sensitivity models are motivated by the logistic model
for the binary indicators $S$ and $R$. Assume the log odds of being
in the trial are linear in $Y(0)$ and $X$ under the logistic selection
specification:
\begin{equation}
P(S=1\mid Y(0),X)=\text{logit}^{-1}\{\alpha_{S}(X)+\gamma_{S}Y(0)\},\label{eq:logistics_selection}
\end{equation}
where $\text{logit}^{-1}(x)=\{1+\exp(-x)\}^{-1}$, and $\alpha_{S}(X)$
can be identified by the observed data once $\gamma_{S}$ is specified.
Using the Bayes rule, the unobserved outcome distribution $f(Y(0)\mid S=1,X)$
is a ``tilting'' version of the observed outcomes:
\begin{align*}
f(Y(0)\mid S=1,X) & =f(Y(0)\mid S=0,X)\frac{P(S=0\mid X)}{P(S=1\mid X)}\times\frac{P(S=1\mid Y(0),X)}{P(S=0\mid Y(0),X)}\\
 & =f(Y(0)\mid S=0,X)\frac{\exp\{\gamma_{S}Y(0)\}}{\E[\exp\{\gamma_{S}Y(0)\}\mid X,S=0]},
\end{align*}
which is free of $\alpha_S(X)$; similar logistic selection specifications can be applied to model the indicator of the intercurrent event within the SAT and external
controls as well. 

\end{remark}

Remark \ref{rmk:logic_selection} implies that our tilting sensitivity model, also known as the exponential tilting model, is connected to the logistic selection specification (\ref{eq:logistics_selection}) with flexible formulation, including many non-parametric
models, such as sieve approximation, Dirichlet process mixtures, and Bayesian additive regression trees. The logistic selection model, despite its drawbacks noted in \cite{copas1997inference}, is widely used to assess selection bias in missing data \citep{robins2000sensitivity, dahabreh2022global} and to conduct sensitivity analyses for unmeasured confounding in causal inference \citep{franks2019flexible, nabi2024semiparametric}.

% Further, the unobserved outcome distribution does not depend on $\alpha_S(X)$, meaning it is not necessary to solve for $\alpha_S(X)$ explicitly.

\subsection{Identification and efficient influence function}

The following theorem establishes the non-parametric identification of
$\tau$ when the sensitivity parameters $\gamma_S$, $\gamma_{R_0}$, and $\gamma_{R_1}$ are fixed. 

\begin{theorem}[Identification under tilting sensitivity models]\label{thm:id_tilting} 

Under Assumptions 1 and 2 in Table \ref{tab:Key-assumptions}, and Model \ref{assump:tilting-MNAR} with fixed $\gamma_{R_{0}}$,
$\gamma_{R_{1}}$ and $\gamma_{S}$, the following identification formula holds for
$\tau$
\begin{align*}
\tau & =\frac{1}{P(S=1)}\E\left[\pi_{S}(X)\pi_{R_{1}}(X)\mu_{1}(X)+\pi_{S}(X)\{1-\pi_{R_{1}}(X)\}\frac{b(X;\gamma_{R_{1}})}{c(X;\gamma_{R_{1}})}\right]\\
 & \ \ -\frac{1}{P(S=1)}\E\left\{ \frac{\pi_{S}(X)d(X;\gamma_{R_{0}},\gamma_{S})}{e(X;\gamma_{R_{0}},\gamma_{S})}\right\} ,
\end{align*}
where the tilted outcomes $b(X;\gamma_{R_{1}})$ and $d(X;\gamma_{R_{0}},\gamma_{S})$,
and the normalizing terms $c(X;\gamma_{R_1})$ and $e(X;\gamma_{R_{0}},\gamma_{S})$ are defined
in Table \ref{tab:Key-assumptions}(B).

\end{theorem}

The identification formula in Theorem \ref{thm:id_tilting} is derived
under the same logic as Theorem \ref{Thm:id_primary}. Under the tilting
sensitivity model, we have 
\begin{align}
 & b(X;\gamma_{R_{1}})/c(X;\gamma_{R_{1}})=\E\{Y(1)\mid X,S=1,R=0\},\label{eq:b/c}\\
 & d(X;\gamma_{R_{0}},\gamma_{S})/e(X;\gamma_{R_{0}},\gamma_{S})=\E\{Y(0)\mid X,S=1\},\label{eq:d/e}
\end{align}
where (\ref{eq:b/c}) reduces to $\mu_{1}(X)$ when $\gamma_{R_{1}}=0$ as the intercurrent events occur at random for SAT,
and (\ref{eq:d/e}) reduces to $\mu_{0}(X)$ when $\gamma_{R_{0}}=\gamma_{S}=0$ as external controls are exchangeable to SAT and the intercurrent
events occur at random for external controls.
Similarly, we can derive the EIF for $\tau$ under Model \ref{assump:tilting-MNAR}
to motivate the semi-parametric efficient estimator.

\begin{theorem}[EIF under tilting sensitivity models]\label{thm:EIF-tilting}
Under the assumptions in Theorem \ref{thm:id_tilting}, the EIF for
$\tau$ with fixed $\gamma_{R_{0}}$,
$\gamma_{R_{1}}$ and $\gamma_{S}$ is
\begin{align}
&\phi_{\mathrm{eff}}^{t}(V;P_{0},\gamma_S, \gamma_{R_{0}},
\gamma_{R_{1}})  =\frac{SRY}{P(S=1)}\label{eq:EIF-observed}\\
 & +\frac{S}{P(S=1)}\left\{(1-R)\frac{b(X;\gamma_{R_{1}})}{c(X;\gamma_{R_{1}})}+Rq_{R_{1}}(X)g(V;\gamma_{R_{1}})\right\}\label{eq:EIF-gamma_R1}\\
 & -\frac{1}{P(S=1)}\left\{ S\frac{d(X;\gamma_{R_{0}},\gamma_{S})}{e(X;\gamma_{R_{0}},\gamma_{S})}+(1-S)q_{S}(X)h(V;\gamma_{R_{0}},\gamma_{S})\right\} -\frac{S\tau}{P(S=1)},\label{eq:EIF-gammaR0-gammaS}
\end{align}
where the augmentation terms $\E\{g(V;\gamma_{R_{1}})\}=0$ and $\E\{h(V;\gamma_{R_{0}},\gamma_{S})\}=0$
with detailed definitions in the Supplementary Materials.

\end{theorem}

The EIF in Theorem \ref{thm:EIF-tilting} is constituted by three
parts. The first part (\ref{eq:EIF-observed}) is contributed by the
SAT with no intercurrent events (i.e., $S=1,R=1$); the second part
(\ref{eq:EIF-gamma_R1}) is contributed by the SAT with intercurrent
events. When $\gamma_{R_{1}}=0$, indicating the intercurrent event
occurs at random for SAT, $g(V;\gamma_{R_{1}})$ equals to $Y-\mu_{1}(X)$,
and the part (\ref{eq:EIF-gamma_R1}) reduces to (\ref{eq:EIF0-part1})
for primary analysis; the third part (\ref{eq:EIF-gammaR0-gammaS})
is contributed by the external controls. When $\gamma_{R_{0}}=\gamma_{S}=0$,
indicating external controls are exchangeable to SAT and the intercurrent
event occurs at random for external controls, $h(V;\gamma_{R_{0}},\gamma_{S})$
equals to $R\{Y-\mu_{0}(X)\}/\pi_{R_{0}}(X)$, and the part (\ref{eq:EIF-gammaR0-gammaS})
reduces to (\ref{eq:EIF0-part2}). Next, we construct an estimator for
$\tau$ by solving the empirical mean of $\phi_{\mathrm{eff}}^{t}(V;P_{0},\gamma_S, \gamma_{R_{0}},
\gamma_{R_{1}})$
with $P_{0}$ replaced by its estimated counterpart, and present the
asymptotic properties for $\widehat{\tau}^{t}$ in Theorem \ref{thm:EIF-tilting-1}.

\begin{theorem}\label{thm:EIF-tilting-1} Under the assumptions in
Theorem \ref{thm:EIF-tilting} and other regularity conditions in
Assumption \ref{assump:regularity}, we have 
\begin{align*}
\widehat{\tau}^{t} & =\frac{1}{N_{\mathcal{R}}}\sum_{i\in\mathcal{R}}R_{i}Y_{i}+\frac{1}{N_{\mathcal{R}}}\sum_{i\in\mathcal{R}}\left[(1-R_{i})\frac{\widehat{b}(X_{i};\gamma_{R_{1}})}{\widehat{c}(X_{i};\gamma_{R_{1}})}+R_{i}\widehat{q}_{R_{1}}(X_{i})\widehat{g}(V_{i};\gamma_{R_{1}})\right]\\
 & \ \ -\frac{1}{N_{\mathcal{R}}}\sum_{i\in\mathcal{R}}\frac{\widehat{d}(X_{i};\gamma_{R_{0}},\gamma_{S})}{\widehat{e}(X_{i};\gamma_{R_{0}},\gamma_{S})}-\frac{1}{N_{\mathcal{R}}}\sum_{i\in\mathcal{\mathcal{E}}}\widehat{q}_{S}(X_{i})\widehat{h}(V_{i};\gamma_{R_{0}},\gamma_{S}).\\
 & =\tau+\frac{1}{N}\sum_{i\in\mathcal{R}\cup\mathcal{E}}\phi_{\mathrm{eff}}^{t}(V_{i};P_{0},\gamma_S, \gamma_{R_{0}},
\gamma_{R_{1}})+\|{\rm Rem}^{t}(\widehat{P},P)\|_{L_{2}}+o_{\pr}(N^{-1/2}),
\end{align*}
where $\|{\rm Rem}^{t}(\widehat{P},P_{0})\|_{L_{2}}$ is bounded by
\begin{align*}
 & \left\{ \|\widehat{q}_{R_{1}}(X)-q_{R_{1}}(X)\|_{L_{2}}+\|\widehat{c}(X;\gamma_{R_{1}})-c(X;\gamma_{R_{1}})\|_{L_{2}}\right\} \\
 & \times\left\{ \|\widehat{c}(X;\gamma_{R_{1}})-c(X;\gamma_{R_{1}})\|_{L_{2}}+\|\widehat{b}(X;\gamma_{R_{1}})-b(X;\gamma_{R_{1}})\|_{L_{2}}\right\} \\
 & +\sum_{\gamma\in\{\gamma_{R_{0}},\gamma_{S},\gamma_{S}+\gamma_{R_{0}}\}}\|\widehat{b}(X;\gamma)-b(X;\gamma)\|_{L_{2}}\\
 & \times\left\{ \|\widehat{q}_{R_{0}}(X)-q_{R_{0}}(X)\|_{L_{2}}+\|\widehat{q}_{S}(X)-q_{S}(X)\|_{L_{2}}+\sum_{\gamma\in\{\gamma_{R_{0}},\gamma_{S},\gamma_{S}+\gamma_{R_{0}}\}}\|\widehat{c}(X;\gamma)-c(X;\gamma)\|_{L_{2}}\right\} \\
 & +\left\{ \|\widehat{q}_{S}(X)-q_{S}(X)\|_{L_{2}}+\sum_{\gamma\in\{\gamma_{R_{0}},\gamma_{S},\gamma_{S}+\gamma_{R_{0}}\}}\|\widehat{c}(X;\gamma)-c(X;\gamma)\|_{L_{2}}\right\} \\
 & \times\left\{ \sum_{\gamma\in\{\gamma_{R_{0}},\gamma_{S},\gamma_{S}+\gamma_{R_{0}}\}}\|\widehat{c}(X;\gamma)-c(X;\gamma)\|_{L_{2}}+\sum_{\gamma\in\{\gamma_{R_{0}},\gamma_{S},\gamma_{S}+\gamma_{R_{0}}\}}\|\widehat{b}(X;\gamma)-b(X;\gamma)\|_{L_{2}}\right\}
\end{align*}
up to some multiplicative constants. 

\end{theorem}

Theorem \ref{thm:EIF-tilting-1} shows that $\widehat{\tau}^{t}$
is root-n consistent and asymptotically normal for fixed sensitivity parameters $\gamma_S$, $\gamma_{R_{0}}$, and $\gamma_{R_{1}}$ when the remainder
term $N^{1/2}\|{\rm Rem}^{t}(\widehat{P},P_{0})\|_{L_{2}}$ is $o_{\pr}(1)$.
Intuitively, when $\gamma_{S}=\gamma_{R_{1}}=\gamma_{R_{0}}=0$, we
have $\|\widehat{c}(X;\gamma)-c(X;\gamma)\|_{L_{2}}=0$ for any $\gamma$,
and
\begin{align*}
\|\widehat{b}(X;\gamma_{R_{1}})-b(X;\gamma_{R_{1}})\|_{L_{2}} & =\|\widehat{\mu}_{1}(X)-\mu_{1}(X)\|_{L_{2}},\\
\sum_{\gamma\in\{\gamma_{R_{0}},\gamma_{S},\gamma_{S}+\gamma_{R_{0}}\}}\|\widehat{b}(X;\gamma)-b(X;\gamma)\|_{L_{2}} & =\|\widehat{\mu}_{0}(X)-\mu_{0}(X)\|_{L_{2}}.
\end{align*}
Thus, the remainder term reduces to:
\begin{align*}
\|{\rm Rem}^{t}(\widehat{P},P_{0})\|_{L_{2}} & \lesssim\|\widehat{q}_{R_{1}}(X)-q_{R_{1}}(X)\|_{L_{2}}\times\|\widehat{\mu}_{1}(X)-\mu_{1}(X)\|_{L_{2}}\\
 & +\|\widehat{\mu}_{0}(X)-\mu_{0}(X)\|_{L_{2}}\times\left\{ \|\widehat{q}_{R_{0}}(X)-q_{R_{0}}(X)\|_{L_{2}}+\|\widehat{q}_{S}(X)-q_{S}(X)\|_{L_{2}}\right\} ,
\end{align*}
which is at the same convergence rate as $\|{\rm Rem}(\widehat{P},P_{0})\|_{L_{2}}$
for the primary analysis in Theorem \ref{Thm:eif0-1}. The remainder term $\|{\rm Rem}^t(\widehat{P},P_0)\|_{L_{2}}$ suggests that the error of $\widehat{\tau}^{[t]}$ is only affected by the estimation errors of the nuisance models in second-order terms. Therefore, $\widehat{\tau}^{[t]}$ is more robust and remains consistent when flexible machine learning methods used for nuisance estimation converge at rates faster than $N^{-1/4}$. This condition is satisfied by some machine learning methods \citep{kennedy2016semiparametric,bradic2019sparsity}, which is the so-called rate double robustness \citep{chernozhukov2018double}. Note that the conditional expectations
$b(X;\gamma)$ and $c(X;\gamma)$ are critical to obtain accurate tilting
estimates for the sensitivity analysis. In general, these conditional
expectations require heavy computation or strong restrictions on the
outcome model. Fortunately, these terms are analytically tractable
when the observed outcome model belongs to the class of exponential
family mixtures, e.g., the Dirichlet processes mixture models \citep{dorie2016flexible}. 

\begin{remark}[Exponential family mixtures]\label{rmk:exp_mixture}

Let the observed outcome follows $Y(0)\mid S=0,R=1,X\sim\sum_{k}\pi_{k}\mathcal{N}(\mu_{0k}(X),\sigma_{0k}^{2}(X))$.
Under the tilting sensitivity models, we can show that $c(X;\gamma_{R_{0}})=\sum_{k}\pi_{k}\exp\{ \mu_{0k}(X)\gamma_{R_{0}}+\gamma_{R_{0}}^{2}\sigma_{0k}^{2}(X)/2\}$, 
and $b(X;\gamma_{R_{0}})=\sum_{k}\pi_{k}\{\mu_{0k}(X)+\gamma_{R_{0}}\sigma_{0k}^{2}(X)\}\exp\{ \mu_{0k}(X)\gamma_{R_{0}}+\gamma_{R_{0}}^{2}\sigma_{0k}^{2}(X)/2\}$. Analogously, these conditional expectations are analytically obtainable
if an invertible function of $Y$ follows the exponential family mixtures
(e.g., box-cox transformation); other advanced methods are also available
to compute the conditional expectations in exchange for heavy computation,
e.g., modeling the conditional distribution of the observed data \citep{chiang2012new}.

\end{remark}

\section{Calibrating sensitivity parameters \label{sec:calibrating}}

The magnitude of the sensitivity parameters indicates the strength
of the non-ignorability of the indicators $(R,S)$ given the covariates, which is commonly caused by the existence of unmeasured confounders. However, it is practically infeasible to identify the sensitivity parameters with the observed data. Furthermore, assessing whether a confounder with such strength plausibly exists, given the prior knowledge and domain expertise, is arguably
challenging as well. While sensitivity parameters are not directly
identifiable from the data, it is reasonable to bound their relative strength using the observed data. 

Following the calibration approach proposed
by \citet{franks2019flexible}, we introduce a method to determine
the plausible quantities for sensitivity analysis based on the observed
data. Assume the logistic selection specification outlined in Remark
\ref{rmk:logic_selection} holds. Next, we assume that the relative strength of the
unmeasured confounder cannot exceed that of the observed covariates,
meaning it should not account for more variation of the indicators
as the most important covariate. To measure the relative strength,
we adopt the ``implicit $R^{2}$'' concept from \citep{imbens2003sensitivity},
which generalizes variance-explained measures to the case of binary
outcomes. For example, the partial variances $\rho_{Y(0)\mid X}^{2}$
explained by $Y(0)$ given $X$ is 
\[
\rho_{Y(0)\mid X}^{2}=\frac{\sigma_{Y}^{2}\gamma_{S}^{2}}{{\rm var}\{m_{S}(X)\}+\pi^{2}/3+\sigma_{Y}^{2}\gamma_{S}^{2}},\quad P(S=1\mid X)=\text{logit}^{-1}\{m_{S}(X)\},
\]
where $\sigma_{Y}^{2}=\E[\text{var}\{Y(0)\mid X,S=0\}]$. Then, we
propose a target value $(\rho^{*})^2$ for the unidentified $\rho_{Y(0)\mid X}^{2}$
using the observed data. In specific, we compute the partial variance
explained by each covariate $X_{j}$ given all other covariates $X_{-j}$,
and set $(\rho^{*})^2=\max_{j}\rho_{X_{j}\mid X_{-j}}^{2}/(1-\max_{j}\rho_{X_{j}\mid X_{-j}}^{2})$. Here, $(\rho^{*})^2$ represents the maximum partial variance explained by adding one covariate
$X_{j}$ to the others, relative to the baseline variance that needs to be explained, referred to as the partial Cohen’s $f$ in \cite{cinelli2020making}. Setting $\rho_{Y(0)\mid X}^{2}=(\rho^{*})^2$
allows us to calibrate $\gamma_{S}$ by $\gamma_{S}^{*}$, which implies
that the information gained by adding $Y(0)$ to $X$ as a predictor
of $S$ is comparable to the maximum information gain by the most
important covariate. To calibrate the sensitivity parameter $\gamma_{S}$,
the following one-to-one mapping is adopted:
\begin{equation}
|\gamma_{S}^{*}|=\frac{1}{\sigma_{Y}}\sqrt{\frac{(\rho^{*})^{2}}{1-(\rho^{*})^{2}}[{\rm var}\{m_{S}(X)\}+\pi^{2}/3]}.\label{eq:gamma_S_calibrated}
\end{equation}
Similar bounding procedures apply to the calibration of $\gamma_{R_{0}}$
and $\gamma_{R_{1}}$.

\section{Simulation study \label{sec:simu}}

\begin{figure}[tbph]
\centering

\includegraphics[width=.8\linewidth]{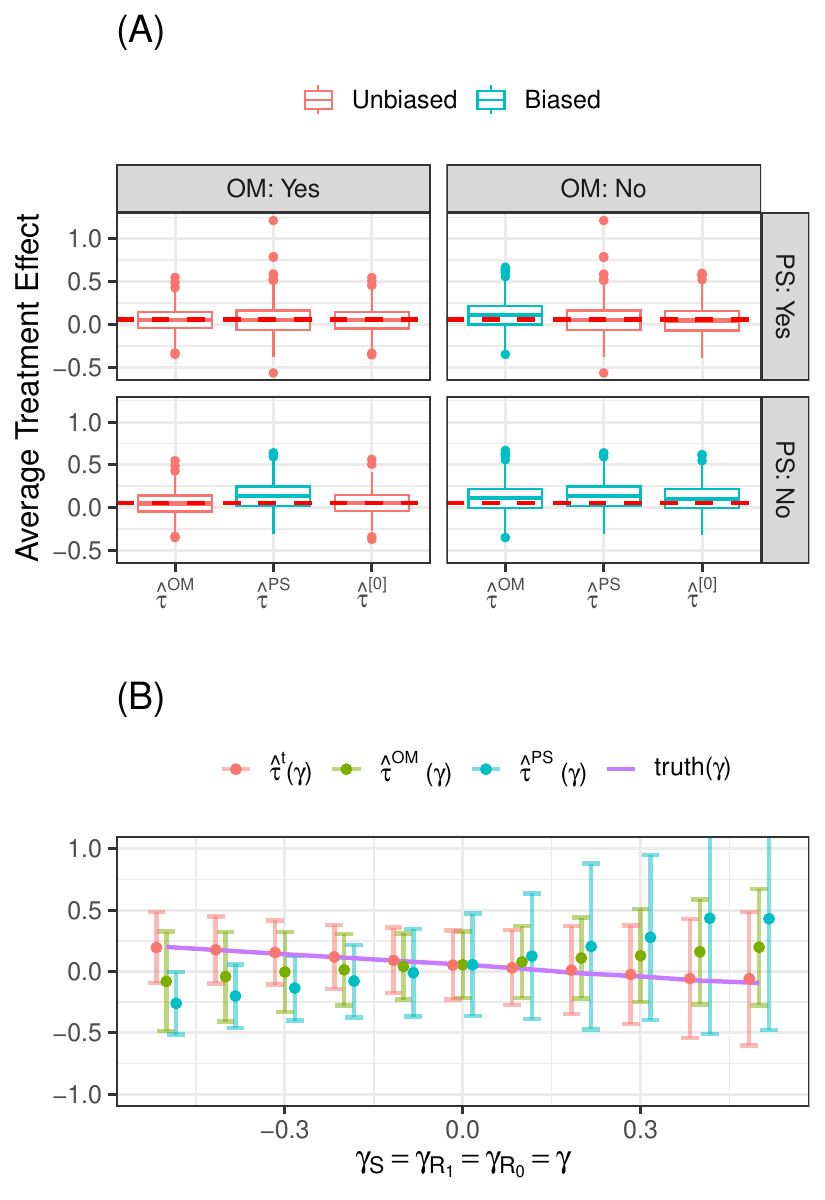}

\caption{\label{fig:sim1} Performance of the PS-based, OR-based, and EIF-motivated tilting estimator with fixed sensitivity parameters: (A) zero-valued under $4$ different
model specifications, and (B) nonzero-valued when all the nuisance models are correctly specified, based on $500$ Monte Carlo simulations.}
\end{figure}

We first conduct a set of simulations to evaluate the operating characteristics
of the proposed estimators under possible model misspecification when
Assumptions 3 to 5 in Table \ref{tab:Key-assumptions} are satisfied.
Set the sample sizes of the SAT and EC to be around $N_{\mathcal{R}}=200$
and $N_{\mathcal{E}}=500$ with total size $N=700$. The covariates
$X\in\mathbb{R}^{5}$ are generated by $X_{j}\sim N(0.25,1)$ for
$j=1,\cdots,4$ and $X_{5}\sim\text{Bernoulli}(0.5)$. Consider a
nonlinear transformation of the covariates and denote $Z_{j}=\{X_{j}^{2}+2\sin(X_{j})-1.5\}/\sqrt{2}$
for $j=1,\cdots,4$ and $Z_{5}=X_{5}$. We generate the indicator
of being selected to SAT or EC by $S\mid X\sim\text{Bernoulli}\{\pi_{S}(X)\},$
where $\pi_{S}(X)=\text{logit}^{-1}(\alpha_{S}+0.1\sum_{j=1}^{5}Z_{j})$
and $\alpha_{S}$ is chosen adaptively to ensure the average of $S$
is about $N_{\mathcal{R}}/N$. Next, we generate the indicators of
the intercurrent events and the outcomes for SAT and EC by 
\begin{align*}
 & R\mid X,S=s\sim\text{Bernoulli}\{\pi_{Rs}(X)\},\quad\pi_{Rs}(X)=\text{logit}^{-1}(\alpha_{R_{S}}+\sum_{j=1}^{5}Z_{j}/6), \quad s=0,1,
\end{align*}
$Y\mid X,S=1\sim\mathcal{N}(\sum_{j=1}^{5}Z_{j}/2,1)$, and $Y\mid X,S=0\sim\mathcal{N}(\sum_{j=1}^{5}Z_{j}/3,1)$, where $(\alpha_{R_{S1}},\alpha_{R_{S0}})$ are adaptively chosen to
ensure the average propensity of $R$ is around $0.5$. With a large
sample size of Monte Carlo simulation, we compute true ATE $\tau=0.13$. First, we assess the robustness of the proposed estimator $\widehat{\tau}^{t}$ when the sensitivity parameters $\gamma_S = \gamma_{R_0}=\gamma_{R_1}=0$. Denote the tilting estimator $\widehat{\tau}^t$ with fixed zero-valued sensitivity parameters as $\widehat{\tau}^{[0]}$, we consider two model specifications of the propensity (PS) of the
participation $\pi_{S}(X)$ and the intercurrent events $\pi_{R_{s}}(X)$,
and the outcome means (OM). In particular, we fit the corresponding
parametric models with the covariates $Z$ as the correctly specified
models or with the covariates $X$ as the misspecified models. We
compare our proposed EIF-motivated tilting estimator with other two estimators,
which are constructed solely based on PS or OM, denoted by $\widehat{\tau}^{PS}$
and $\widehat{\tau}^{OM}$, respectively. Figure \ref{fig:sim1}(A) shows
the point estimation results based on $500$ Monte Carlo experiments.
When PS and OM are correctly specified, the considered estimators
are all unbiased. However, $\widehat{\tau}^{PS}$ and $\widehat{\tau}^{OM}$
are biased when their required models are misspecified. Our proposed
tilting estimator is shown to be doubly robust with fixed zero-valued sensitivity parameters as it is consistent if either
PS or OM is correctly specified.

Table \ref{tab:sim1}(A) shows the absolute bias, standard errors (SE), mean
squared error (MSE), coverage rates (CR), and the average CI lengths
of each estimator. We construct the corresponding $95\%$ Wald-type
CIs for inference, where the variances are estimated by non-parametric
bootstrap with size $B=50$. We observe that both $\widehat{\tau}^{[0]}$ and $\widehat{\tau}^{OM}$ exhibit the smallest average confidence interval lengths. However, when the outcome model (OM) is misspecified, the coverage rate for $\widehat{\tau}^{[0]}$ is closer to the nominal level compared to $\widehat{\tau}^{OM}$. This observation underscores the double robustness of $\widehat{\tau}^{[0]}$, aligning with the results shown in Figure \ref{fig:sim1} and supporting our claims in Theorem \ref{Thm:eif0-1} for the primary analysis.

Next, we assess the performance of our proposed tilting estimators
$\widehat{\tau}^{t}$ when Assumptions 3 to 5 in Table \ref{tab:Key-assumptions}
are violated. We keep the same data-generating process for the $X$,
$Z$ and $Y$ but generate the indicators using Bernoulli sampling
with different propensities:
\begin{align*}
P\{S=1 & \mid X,Y(0)\}=\text{logit}^{-1}(\alpha_{S}+0.1\sum_{j=1}^{5}Z_{j}+\gamma_{S}Y(0)),\\
P\{R=0 & \mid X,Y(s),S=s\}=\text{logit}^{-1}(-\alpha_{R_{s}}-\sum_{j=1}^{5}Z_{j}/6+\gamma_{R_{s}}Y(s)),\quad s=0,1,
\end{align*}
where $(\gamma_{S},\gamma_{R_{1}},\gamma_{R_{0}})$ control the strength
of the unmeasured confounder, describing how the propensity depends
on the potential outcome after accounting for the covariates. We consider
a range of sensitivity parameters where the indicators are equally
confounded by the potential outcomes, i.e., $\gamma=\gamma_{S}=\gamma_{R_{1}}=\gamma_{R_{0}}$. Under the hypothetically fixed sensitivity parameters
$(\gamma_{S},\gamma_{R_{1}},\gamma_{R_{0}})$ and assuming that all the nuisance models are correctly specified, Figure \ref{fig:sim1}(B) shows the point estimation for all three estimators, and Table \ref{tab:sim1}(B) presents the finite-sample performances
of $\widehat{\tau}^{t}$ in details. One important finding is that our proposed estimator is always consistent across a range of fixed sensitivity parameters, as indicated by its small bias and approximately correct coverage rates. In contrast, the other two estimators are unable to handle the joint sensitivity analysis for multiple assumptions, suggested by their non-negligible biases for fixed non-zero sensitivity parameters.

\begin{table}[tbph]
\centering
\caption{\label{tab:sim1} The bias, standard errors (SE), mean squared error
(MSE), coverage rates (CR), and the average CI width based on $500$
Monte Carlo experiments of (A) the PS-based, OR-based, and EIF-motivated
estimator under $4$ different model specifications when Assumptions
3 to 5 in Table \ref{tab:Key-assumptions} hold; (B) the EIF-motivated
tilting estimates with the fixed sensitivity parameters when Assumptions
3 to 5 in Table \ref{tab:Key-assumptions} are violated}
\begin{tabular}{lccccccc}
\hline 
(A)&
\multicolumn{2}{c}{} & bias & SE & MSE & CR & CI width\tabularnewline
\hline 
\multicolumn{8}{c}{PS-based estimator $\widehat{\tau}^{PS}$}\tabularnewline
\hline 
&PS = yes & OM = yes & 0.01 & 0.22 & 0.05 & 95.8\% & 0.86\tabularnewline
&PS = yes & OM = no & 0.01 & 0.22 & 0.05 & 95.8\% & 0.86\tabularnewline
&PS = no & OM = yes & 0.07 & 0.18 & 0.04 & 91.4\% & 0.86\tabularnewline
&PS = no & OM = no & 0.07 & 0.18 & 0.04 & 86.4\% & 0.86\tabularnewline
\hline 
\multicolumn{8}{c}{OM-based estimator $\widehat{\tau}^{OM}$}\tabularnewline
\hline 
&PS = yes & OM = yes & 0.02 & 0.14 & 0.02 & 95.0\% & 0.55\tabularnewline
&PS = yes & OM = no & 0.05 & 0.17 & 0.03 & 86.4\% & 0.55\tabularnewline
&PS = no & OM = yes & 0.02 & 0.14 & 0.02 & 95.0\% & 0.55\tabularnewline
&PS = no & OM = no & 0.05 & 0.17 & 0.03 & 86.4\% & 0.55\tabularnewline
\hline 
\multicolumn{8}{c}{EIF-motivated tilting estimator $\widehat{\tau}^{[0]}$}\tabularnewline
\hline 
&PS = yes & OM = yes & 0.02 & 0.14 & 0.02 & 94.2\% & 0.56\tabularnewline
&PS = yes & OM = no & 0.02 & 0.18 & 0.03 & 93.2\% & 0.56\tabularnewline
&PS = no & OM = yes & 0.02 & 0.14 & 0.02 & 95.2\% & 0.56\tabularnewline
&PS = no & OM = no & 0.04 & 0.17 & 0.03 & 89.8\% & 0.56\tabularnewline
\bottomrule 
\\
\toprule
(B)&
\multicolumn{2}{c}{$\gamma_{S}=\gamma_{R_{1}}=\gamma_{R_{0}}$} & bias & SE & MSE & CR & CI width\tabularnewline
\hline 
\multicolumn{8}{c}{EIF-motivated tilting estimator $\widehat{\tau}^{t}$}\tabularnewline
\hline 
&\multicolumn{2}{c}{$-0.5$} & 0.01 & 0.15 & 0.02 & 95.2\% & 0.58\tabularnewline
&\multicolumn{2}{c}{$-0.3$} & 0.01 & 0.13 & 0.02 & 95.4\% & 0.52\tabularnewline
&\multicolumn{2}{c}{$-0.1$} & 0.01 & 0.14 & 0.02 & 96.2\% & 0.54\tabularnewline
&\multicolumn{2}{c}{$0.0$} & 0.00 & 0.14 & 0.02 & 94.2\% & 0.56\tabularnewline
&\multicolumn{2}{c}{$0.1$} & 0.01 & 0.16 & 0.02 & 94.8\% & 0.61\tabularnewline
&\multicolumn{2}{c}{$0.3$} & 0.02 & 0.22 & 0.05 & 93.6\% & 0.80\tabularnewline
&\multicolumn{2}{c}{$0.5$} & 0.03 & 0.31 & 0.09 & 93.4\% & 1.09\tabularnewline
\hline 
\end{tabular}
\end{table}

\section{Real-data application}\label{sec:real}

We examine a study designed to estimate an antidepressant drug effect
on the scores of the Hamilton Depression Rating Scale for 17 items (HAMD-17).
This study was conducted under the Auspices of the Drug Information
Association, which collects the data at baseline and weeks 1, 2, 4,
6, and 8 for $N=196$ patients with $99$ in the control group and
$97$ in the treatment group. However,
some patients may drop out during the study for various reasons. Our
primary interest is the ATE on the change of HAMD-17, irrespective of the intercurrent
events such as dropout-related missing data. According to the guidelines
in \citet{international2019addendum}, the ATE is defined as the mean
difference of the change in the HAMD-17 scores from the baseline to
the final time point week $8$. We adhere to the same analysis plan
as \citet{liu2024multiply} with covariates $X$ including the investigation
sites and baseline HAMD-17 scores. Let $Y(a)$ and $R$ be the change
of HAMD-17 scores under treatment $a$ and the indicator of whether a patient stayed in the study at week 8.

\begin{figure}[htbp]
\centering

\includegraphics[width=0.8\linewidth]{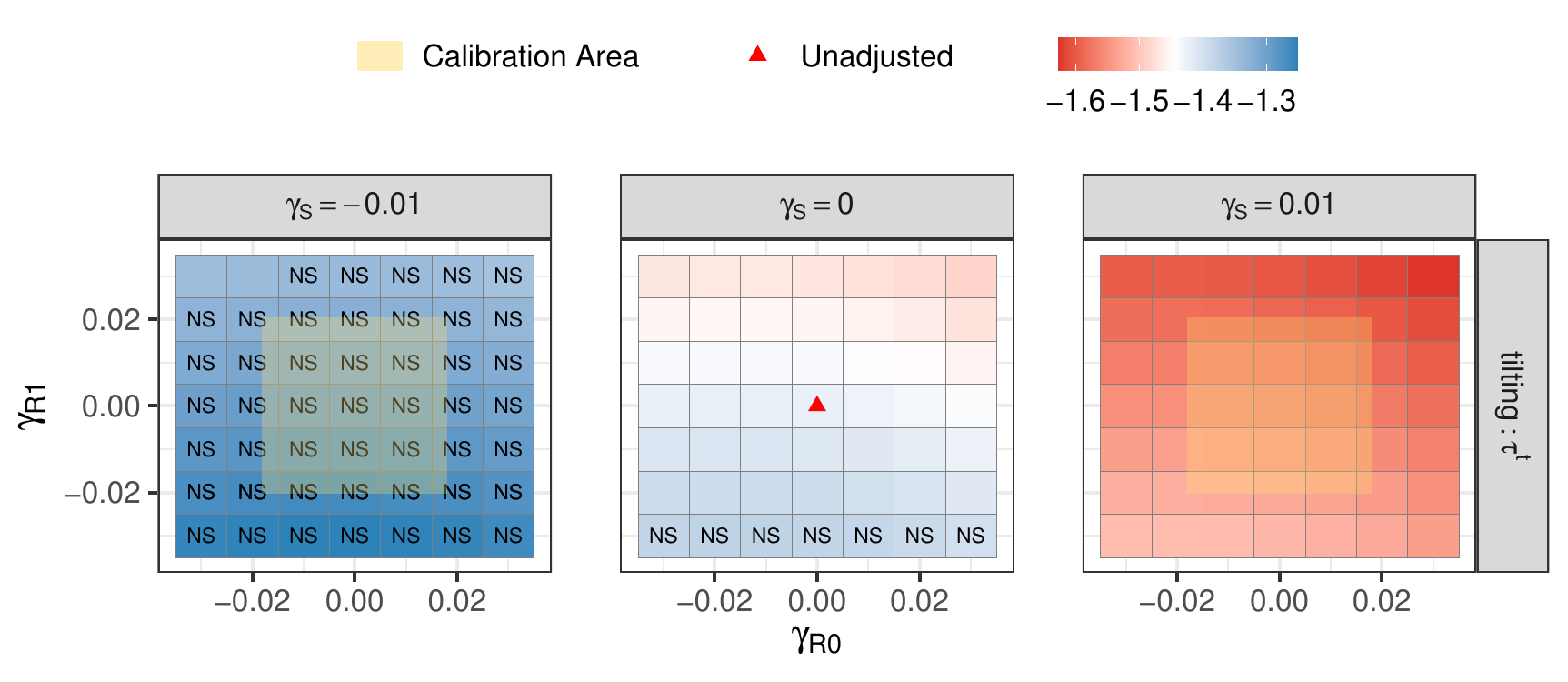}\caption{\label{fig:ATE_grid} Average treatment effects of an antidepressant
drug effect on the HAMD-17 scores over a grid of hypothesized sensitivity
parameters under the tilting sensitivity models. }

\end{figure}

In our application, we consider the original trial as a single-arm
trial, with the concurrent control group being considered as external
controls to illustrate the proposed sensitivity analysis.  We assume the potential outcomes $Y(a)$ follow a Gaussian
mixture model fitted using the R package \texttt{flexmix}. Next, we bound the
magnitude of the sensitivity parameters using the approach described
in Section \ref{sec:calibrating}. We illustrate this approach with
the baseline HAMD-17 score, which is the most important predictors
in terms of partial variance explained, with $(\rho_{S}^{*})^2\approx0.02$,
$(\rho_{R_{1}}^{*})^2\approx0.11$ and $(\rho_{R_{0}}^{*})^2\approx0.04$.
To map these values to the sensitivity parameters, we apply the one-to-one
mapping formula (\ref{eq:gamma_S_calibrated}), and obtain the calibrated
sensitivity parameters $|\gamma_{R_{0}}^{*}|\approx0.02$, $|\gamma_{R_{1}}^{*}|\approx0.02$,
and $|\gamma_{S}^{*}|\approx0.01$. Figure \ref{fig:ATE_grid} illustrates
the ATE estimates across a range of hypothetical sensitivity parameters, adjusting for the potential outcomes as the unmeasured confounder
in the logistic selection specification. The shaded area indicates
the unmeasured confounder with impacts up to the values of the calibrated
sensitivity parameters. Here, \textquotedblleft NS\textquotedblright{}
denotes \textquotedblleft not significant,\textquotedblright{} meaning
the $95\%$ confidence interval of the ATE contains $0$. 

When the assumptions are satisfied, that is, all the sensitivity parameters equal zero, the ATE estimates are $\widehat{\tau}^{t} = -1.42$ with the $95\%$ bootstrap CI as $(-2.80, -0.05)$, which is statistically significant. Next, we assume the impact of the confounders act towards
hurting our preferred hypothesis, that is, $\gamma_{S}^{*}<0$.
Here, a negative value of $\gamma_{S}$ suggests that the unobserved
change in HAMD-17 scores in the concurrent control group tend to be
lower (i.e., better) than the observed external controls, reducing
the absolute value of the effect size. This could occur if patients
are more likely to participate in the single-arm trial when less depressed.
When $\gamma_{S} = -0.01$ and $\gamma_{R_{0}} = \gamma_{R_{1}} = -0.02$, the estimated treatment effect of the antidepressant drug becomes $\widehat{\tau}^{t} = -1.28$, where the unmeasured confounder is as strong as the baseline HAMD-17 scores. Although the tilted estimate is below zero, suggesting the effectiveness of the drug on the HAMD-17 scores, it is no longer statistically significant, as its $95\%$ CI is $(-2.65, 0.09)$. Thus, following our sensitivity analyses, we show that the magnitude of the 
possible drug effects on the HAMD-17 scores is robust but the significance of such effect is not robust against unmeasured confounding at the strength of baseline HAMD-17 scores. However, domain knowledge is still required to consider the plausibility of an unmeasured confounders of such strength level under this situation.

\section{Discussion}\label{sec:discuss}

In this paper, we develop a semi-parametric efficient framework for sensitivity analysis under the tilting models. Motivated by Tukey’s factorization, this framework effectively separates the model checking from the sensitivity analysis, which does not rely on any modeling assumption and fits perfectly well with the EIF-motivated tilting estimators. Decoupling the sensitivity analysis from the model fit assessment is crucial and ubiquitous within the model-based sensitivity analysis. However, joint sensitivity analysis for multiple assumptions remains largely unexplored to the best of our knowledge. By 
simultaneously assessing the EC outcome mean non-exchangeability and the effects of intercurrent events, our framework hopes to shed more light on the advancements of joint modeling for sensitivity analysis.

Future work could extend our framework to longitudinal trials with intercurrent events, particularly those with irregular and informative observation patterns, as discussed by \cite{yang2021semiparametric} and \cite{smith2022trials}. Such an extension may increase the number of sensitivity parameters across multiple time points, and introduce additional challenges in deriving the EIFs conditioned on the historical information. Another potential extension could focus on the choices of sensitivity parameters, which is profoundly useful in practice. Our approach relies on bounding the magnitude of sensitivity parameters using observed data, and substantive domain expertise should be consulted to examine whether an unmeasured confounder with such strength is plausible. For some hybrid control designs, the sensitivity parameters can be partially identified with the help of concurrent controls; similar ideas have been explored in \cite{gao2023integrating} to adjust for EC outcome mean non-exchangeability. Thus, the internal validity from the hybrid controls can be leveraged to inform the choices of sensitivity parameters. In summary, our proposed semi-parametric sensitivity analysis is both efficient and flexible as it is rate-doubly robust, locally optimal, and
can be incorporated with a range of models with modern causal inference workflow.

\section*{Supplementary Materials}
Web Appendices, tables, figures, and the proofs of all the theorems referenced in Sections \ref{sec:1time}, \ref{sec:Sensitivity-analysis}, and \ref{sec:calibrating}, along with additional details of the J2R sensitivity model, are available with this paper on the Biometrics website at Oxford Academic. An R-package for implementing the proposed method is available at \url{https://github.com/Gaochenyin/SensDR}. 

\section*{Acknowledgments}
We thank the editor, associate editor, two anonymous referees and Yujing Gao from North Carolina State University for their constructive suggestions and valuable feedback on our manuscript.

\section*{Data Availability}
The data that support the findings in this paper are available in the Drug Information Association Missing Data at \url{https://www.lshtm.ac.uk/research/centres-projects-groups/missing-data#dia-missing-data}, collected by \cite{mallinckrodt2014recent}, and are also provided at \url{https://github.com/Gaochenyin/SensDR}.

\bibliographystyle{biom}
\bibliography{ci}

\begin{thebibliography}{}

\bibitem[\protect\citeauthoryear{Allen, Satten, and Tsiatis}{Allen et~al.}{2005}]{allen2005locally}
Allen, A.~S., Satten, G.~A., and Tsiatis, A.~A. (2005).
\newblock Locally-efficient robust estimation of haplotype-disease association in family-based studies.
\newblock {\em Biometrika} {\bf 92,} 559--571.

\bibitem[\protect\citeauthoryear{Blackwell}{Blackwell}{2014}]{blackwell2014selection}
Blackwell, M. (2014).
\newblock A selection bias approach to sensitivity analysis for causal effects.
\newblock {\em Political Analysis} {\bf 22,} 169--182.

\bibitem[\protect\citeauthoryear{Bradic, Wager, and Zhu}{Bradic et~al.}{2019}]{bradic2019sparsity}
Bradic, J., Wager, S., and Zhu, Y. (2019).
\newblock Sparsity double robust inference of average treatment effects.
\newblock {\em arXiv:1905.00744} .

\bibitem[\protect\citeauthoryear{Carpenter, Roger, and Kenward}{Carpenter et~al.}{2013}]{carpenter2013analysis}
Carpenter, J.~R., Roger, J.~H., and Kenward, M.~G. (2013).
\newblock Analysis of longitudinal trials with protocol deviation: a framework for relevant, accessible assumptions, and inference via multiple imputation.
\newblock {\em Journal of Biopharmaceutical Statistics} {\bf 23,} 1352--1371.

\bibitem[\protect\citeauthoryear{Chernozhukov, Chetverikov, Demirer, Duflo, Hansen, Newey, and Robins}{Chernozhukov et~al.}{2018}]{chernozhukov2018double}
Chernozhukov, V., Chetverikov, D., Demirer, M., Duflo, E., Hansen, C., Newey, W., and Robins, J. (2018).
\newblock Double/debiased machine learning for treatment and structural parameters.
\newblock {\em Econometrics Journal} {\bf 21,} C1--C68.

\bibitem[\protect\citeauthoryear{Chiang and Huang}{Chiang and Huang}{2012}]{chiang2012new}
Chiang, C.-T. and Huang, M.-Y. (2012).
\newblock New estimation and inference procedures for a single-index conditional distribution model.
\newblock {\em Journal of Multivariate Analysis} {\bf 111,} 271--285.

\bibitem[\protect\citeauthoryear{Cinelli and Hazlett}{Cinelli and Hazlett}{2020}]{cinelli2020making}
Cinelli, C. and Hazlett, C. (2020).
\newblock Making sense of sensitivity: Extending omitted variable bias.
\newblock {\em Journal of the Royal Statistical Society: Series B (Statistical Methodology)} {\bf 82,} 39--67.

\bibitem[\protect\citeauthoryear{Copas and Li}{Copas and Li}{1997}]{copas1997inference}
Copas, J.~B. and Li, H. (1997).
\newblock Inference for non-random samples.
\newblock {\em Journal of the Royal Statistical Society Series B: Statistical Methodology} {\bf 59,} 55--95.

\bibitem[\protect\citeauthoryear{Cornfield, Haenszel, Hammond, Lilienfeld, Shimkin, and Wynder}{Cornfield et~al.}{1959}]{cornfield1959smoking}
Cornfield, J., Haenszel, W., Hammond, E.~C., Lilienfeld, A.~M., Shimkin, M.~B., and Wynder, E.~L. (1959).
\newblock Smoking and lung cancer: recent evidence and a discussion of some questions.
\newblock {\em Journal of the National Cancer Institute} {\bf 22,} 173--203.

\bibitem[\protect\citeauthoryear{Dahabreh, Robins, Haneuse, Saeed, Robertson, Stuart, and Hern{\'a}n}{Dahabreh et~al.}{2023}]{dahabreh2022global}
Dahabreh, I.~J., Robins, J.~M., Haneuse, S. J.-P., Saeed, I., Robertson, S.~E., Stuart, E.~A., and Hern{\'a}n, M.~A. (2023).
\newblock Sensitivity analysis using bias functions for studies extending inferences from a randomized trial to a target population.
\newblock {\em Statistics in Medicine} {\bf 42,} 2029--2043.

\bibitem[\protect\citeauthoryear{Dorie, Harada, Carnegie, and Hill}{Dorie et~al.}{2016}]{dorie2016flexible}
Dorie, V., Harada, M., Carnegie, N.~B., and Hill, J. (2016).
\newblock A flexible, interpretable framework for assessing sensitivity to unmeasured confounding.
\newblock {\em Statistics in medicine} {\bf 35,} 3453--3470.

\bibitem[\protect\citeauthoryear{Faries, Gao, Zhang, Hazlett, Stamey, Yang, Ding, Shan, Sheffield, and Dreyer}{Faries et~al.}{2024}]{faries2023real}
Faries, D., Gao, C., Zhang, X., Hazlett, C., Stamey, J., Yang, S., Ding, P., Shan, M., Sheffield, K., and Dreyer, N. (2024).
\newblock Real effect or bias? best practices for evaluating the robustness of real-world evidence through quantitative sensitivity analysis for unmeasured confounding.
\newblock {\em Pharmaceutical Statistics} .

\bibitem[\protect\citeauthoryear{{Food and Drug Administration}}{{Food and Drug Administration}}{2023}]{FDA2023}
{Food and Drug Administration} (2023).
\newblock Considerations for the design and conduct of externally controlled trials for drug and biological products guidance for industry.
\newblock \url{https://www.fda.gov/media/164960/download}.
\newblock Accessed: 2023-02-23.

\bibitem[\protect\citeauthoryear{Franks, D'Amour, and Feller}{Franks et~al.}{2020}]{franks2019flexible}
Franks, A., D'Amour, A., and Feller, A. (2020).
\newblock Flexible sensitivity analysis for observational studies without observable implications.
\newblock {\em Journal of the American Statistical Association} {\bf 115,} 1730--1746.

\bibitem[\protect\citeauthoryear{Gao, Yang, Shan, YE, Lipkovich, and Faries}{Gao et~al.}{2024}]{gao2023integrating}
Gao, C., Yang, S., Shan, M., YE, W., Lipkovich, I., and Faries, D. (2024).
\newblock Improving randomized controlled trial analysis via data-adaptive borrowing.
\newblock {\em Biometrika} page asae069.

\bibitem[\protect\citeauthoryear{Garcia and Ma}{Garcia and Ma}{2016}]{garcia2016optimal}
Garcia, T.~P. and Ma, Y. (2016).
\newblock Optimal estimator for logistic model with distribution-free random intercept.
\newblock {\em Scandinavian Journal of Statistics} {\bf 43,} 156--171.

\bibitem[\protect\citeauthoryear{Heckman}{Heckman}{1979}]{heckman1979sample}
Heckman, J.~J. (1979).
\newblock Sample selection bias as a specification error.
\newblock {\em Econometrica: Journal of the econometric society} pages 153--161.

\bibitem[\protect\citeauthoryear{ICH}{ICH}{2021}]{international2019addendum}
ICH (2021).
\newblock {E9(R1)} statistical principles for clinical trials: Addendum: Estimands and sensitivity analysis in clinical trials.
\newblock {\em FDA Guidance Documents} .

\bibitem[\protect\citeauthoryear{Imbens}{Imbens}{2003}]{imbens2003sensitivity}
Imbens, G.~W. (2003).
\newblock Sensitivity to exogeneity assumptions in program evaluation.
\newblock {\em American Economic Review} {\bf 93,} 126--132.

\bibitem[\protect\citeauthoryear{Imbens and Rubin}{Imbens and Rubin}{2015}]{imbens2015causal}
Imbens, G.~W. and Rubin, D.~B. (2015).
\newblock {\em Causal inference in statistics, social, and biomedical sciences}.
\newblock Cambridge University Press.

\bibitem[\protect\citeauthoryear{Kennedy}{Kennedy}{2016}]{kennedy2016semiparametric}
Kennedy, E.~H. (2016).
\newblock Semiparametric theory and empirical processes in causal inference.
\newblock In {\em Statistical causal inferences and their applications in public health research}, pages 141--167. Springer.

\bibitem[\protect\citeauthoryear{Lipkovich, Ratitch, and Mallinckrodt}{Lipkovich et~al.}{2020}]{lipkovich2020causal}
Lipkovich, I., Ratitch, B., and Mallinckrodt, C.~H. (2020).
\newblock Causal inference and estimands in clinical trials.
\newblock {\em Statistics in Biopharmaceutical Research} {\bf 12,} 54--67.

\bibitem[\protect\citeauthoryear{Little and Yau}{Little and Yau}{1996}]{little1996intent}
Little, R. and Yau, L. (1996).
\newblock Intent-to-treat analysis for longitudinal studies with drop-outs.
\newblock {\em Biometrics} pages 1324--1333.

\bibitem[\protect\citeauthoryear{Little and Rubin}{Little and Rubin}{2019}]{little2019statistical}
Little, R.~J. and Rubin, D.~B. (2019).
\newblock {\em Statistical analysis with missing data}, volume 793.
\newblock John Wiley \& Sons.

\bibitem[\protect\citeauthoryear{Liu, Yang, Zhang, and Liu}{Liu et~al.}{2024}]{liu2024multiply}
Liu, S., Yang, S., Zhang, Y., and Liu, G. (2024).
\newblock Multiply robust estimators in longitudinal studies with missing data under control-based imputation.
\newblock {\em Biometrics} {\bf 80,} ujad036.

\bibitem[\protect\citeauthoryear{Liu, Zhang, Golm, Liu, and Yang}{Liu et~al.}{2024}]{liu2024robust}
Liu, S., Zhang, Y., Golm, G.~T., Liu, G., and Yang, S. (2024).
\newblock Robust analyzes for longitudinal clinical trials with missing and non-normal continuous outcomes.
\newblock {\em Statistical theory and related fields} {\bf 8,} 1--14.

\bibitem[\protect\citeauthoryear{Mallinckrodt, Roger, Chuang-Stein, Molenberghs, O'Kelly, Ratitch, Janssens, and Bunouf}{Mallinckrodt et~al.}{2014}]{mallinckrodt2014recent}
Mallinckrodt, C., Roger, J., Chuang-Stein, C., Molenberghs, G., O'Kelly, M., Ratitch, B., Janssens, M., and Bunouf, P. (2014).
\newblock Recent developments in the prevention and treatment of missing data.
\newblock {\em Therapeutic innovation \& regulatory science} {\bf 48,} 68--80.

\bibitem[\protect\citeauthoryear{Nabi, Bonvini, Kennedy, Huang, Smid, and Scharfstein}{Nabi et~al.}{2024}]{nabi2024semiparametric}
Nabi, R., Bonvini, M., Kennedy, E.~H., Huang, M.-Y., Smid, M., and Scharfstein, D.~O. (2024).
\newblock Semiparametric sensitivity analysis: unmeasured confounding in observational studies.
\newblock {\em Biometrics} {\bf 80,} ujae106.

\bibitem[\protect\citeauthoryear{Robins, Rotnitzky, and Scharfstein}{Robins et~al.}{2000}]{robins2000sensitivity}
Robins, J.~M., Rotnitzky, A., and Scharfstein, D.~O. (2000).
\newblock Sensitivity analysis for selection bias and unmeasured confounding in missing data and causal inference models.
\newblock In {\em Statistical models in epidemiology, the environment, and clinical trials}, pages 1--94. Springer.

\bibitem[\protect\citeauthoryear{Rosenbaum}{Rosenbaum}{1987}]{rosenbaum1987sensitivity}
Rosenbaum, P.~R. (1987).
\newblock Sensitivity analysis for certain permutation inferences in matched observational studies.
\newblock {\em Biometrika} {\bf 74,} 13--26.

\bibitem[\protect\citeauthoryear{Rosenbaum and Rubin}{Rosenbaum and Rubin}{1983a}]{rosenbaum1983assessing}
Rosenbaum, P.~R. and Rubin, D.~B. (1983a).
\newblock Assessing sensitivity to an unobserved binary covariate in an observational study with binary outcome.
\newblock {\em Journal of the Royal Statistical Society: Series B (Statistical Methodology)} {\bf 45,} 212--218.

\bibitem[\protect\citeauthoryear{Rosenbaum and Rubin}{Rosenbaum and Rubin}{1983b}]{rosenbaum1983central}
Rosenbaum, P.~R. and Rubin, D.~B. (1983b).
\newblock The central role of the propensity score in observational studies for causal effects.
\newblock {\em Biometrika} {\bf 70,} 41--55.

\bibitem[\protect\citeauthoryear{Smith, Gao, Yang, Varadhan, Apter, and Scharfstein}{Smith et~al.}{2024}]{smith2022trials}
Smith, B.~B., Gao, Y., Yang, S., Varadhan, R., Apter, A.~J., and Scharfstein, D.~O. (2024).
\newblock Semi-parametric sensitivity analysis for trials with irregular and informative assessment times.
\newblock {\em Biometrics} {\bf 80,} ujae154.

\bibitem[\protect\citeauthoryear{Tan, Cro, Van~Vogt, Szigeti, and Cornelius}{Tan et~al.}{2021}]{tan2021review}
Tan, P.-T., Cro, S., Van~Vogt, E., Szigeti, M., and Cornelius, V.~R. (2021).
\newblock A review of the use of controlled multiple imputation in randomised controlled trials with missing outcome data.
\newblock {\em BMC Medical Research Methodology} {\bf 21,} 1--17.

\bibitem[\protect\citeauthoryear{Tsiatis}{Tsiatis}{2006}]{tsiatis2006semiparametric}
Tsiatis, A.~A. (2006).
\newblock {\em Semiparametric theory and missing data}, volume~4.
\newblock Springer.

\bibitem[\protect\citeauthoryear{Tsiatis and Ma}{Tsiatis and Ma}{2004}]{tsiatis2004locally}
Tsiatis, A.~A. and Ma, Y. (2004).
\newblock Locally efficient semiparametric estimators for functional measurement error models.
\newblock {\em Biometrika} {\bf 91,} 835--848.

\bibitem[\protect\citeauthoryear{VanderWeele and Arah}{VanderWeele and Arah}{2011}]{vanderweele2011bias}
VanderWeele, T.~J. and Arah, O.~A. (2011).
\newblock Bias formulas for sensitivity analysis of unmeasured confounding for general outcomes, treatments, and confounders.
\newblock {\em Epidemiology} pages 42--52.

\bibitem[\protect\citeauthoryear{Veitch and Zaveri}{Veitch and Zaveri}{2020}]{veitch2020sense}
Veitch, V. and Zaveri, A. (2020).
\newblock Sense and sensitivity analysis: Simple post-hoc analysis of bias due to unobserved confounding.
\newblock {\em Advances in Neural Information Processing Systems} {\bf 33,} 10999--11009.

\bibitem[\protect\citeauthoryear{Wellner et~al\mbox{.}}{Wellner et~al.}{2013}]{wellner2013weak}
Wellner, J. et~al. (2013).
\newblock {\em Weak convergence and empirical processes: with applications to statistics}.
\newblock Springer Science \& Business Media.

\bibitem[\protect\citeauthoryear{Yang}{Yang}{2021}]{yang2021semiparametric}
Yang, S. (2021).
\newblock {Semiparametric Estimation of Structural Nested Mean Models with Irregularly Spaced Longitudinal Observations}.
\newblock {\em Biometrics} {\bf 78,} 937--949.

\bibitem[\protect\citeauthoryear{Yang and Lok}{Yang and Lok}{2018}]{yang2018sensitivity}
Yang, S. and Lok, J.~J. (2018).
\newblock Sensitivity analysis for unmeasured confounding in coarse structural nested mean models.
\newblock {\em Statistica Sinica} {\bf 28,} 1703.

\bibitem[\protect\citeauthoryear{Yang, Zhang, Liu, and Guan}{Yang et~al.}{2023}]{yang2023smim}
Yang, S., Zhang, Y., Liu, G.~F., and Guan, Q. (2023).
\newblock Smim: a unified framework of survival sensitivity analysis using multiple imputation and martingale.
\newblock {\em Biometrics} {\bf 79,} 230--240.

\bibitem[\protect\citeauthoryear{Zhang and Tchetgen}{Zhang and Tchetgen}{2019}]{zhang2019semiparametric}
Zhang, B. and Tchetgen, E. J.~T. (2019).
\newblock A semiparametric approach to model-based sensitivity analysis in observational studies.
\newblock {\em arXiv preprint arXiv:1910.14130} .

\end{thebibliography}

\newpage
\section*{Appendix}
\section{Proofs}
\subsection{Proof of Theorem \ref{Thm:id_primary}}

Here, we show that $\tau$ can be identified by the first identification
formula in Theorem \ref{Thm:id_primary}.
\begin{align*}
\tau & =\E\{Y\{1,R(1)\}-Y\{0,R(0)\}\mid S=1\}.
\end{align*}
For the term $\E[Y\{1,R(1)\}\mid S=1]$, we have 
\begin{align*}
& \E[Y\{1,R(1)\}\mid S=1] \\
&=\frac{1}{P(S=1)}\E[SY\{1,R(1)\}]\\
 & =\frac{1}{P(S=1)}\E[SR(1)Y(1,1)+S\{1-R(1)\}Y(1,0)]\\
 & =\frac{1}{P(S=1)}\E\left(\E[SR(1)Y(1,1)+S\{1-R(1)\}Y(1,0)\mid X]\right)\\
 & =\frac{1}{P(S=1)}\E\left[\pi_{S}(X)\E\{R(1)\mid X,S=1\}\E\{Y(1,1)\mid X,R(1)=1,S=1\}\right]\\
 & +\frac{1}{P(S=1)}\E\left[\pi_{S}(X)\E\{1-R(1)\mid X,S=1\}\E\{Y(1,0)\mid X,R(1)=0,S=1\}\right]\\
 & =\frac{1}{P(S=1)}\E\{\pi_{S}(X)\pi_{R_{1}}(X)\mu_{1}(X)\}+\frac{1}{P(S=1)}\E\left[\pi_{S}(X)\{1-\pi_{R_{1}}(X)\}\mu_{1}(X)\right]\\
 & =\frac{1}{P(S=1)}\E\{\pi_{S}(X)\mu_{1}(X)\},
\end{align*}
where the equality holds under the assumptions $R=R(A),$ $Y=Y\left\{ A,R(A)\right\} $,
and $R(a)\perp Y(a,r)\mid X,S=1$. Similar technique gives us $\E[Y\{0,R(0)\}\mid S=1]=\pi_{S}(X)\mu_{0}(X)\}/P(S=1)$.
Putting these two terms together gives us:
\[
\tau=\frac{1}{P(S=1)}\E\{\pi_{S}(X)\mu_{1}(X)-\pi_{S}(X)\mu_{0}(X)\},
\]
which proves the first identification formula in Theorem \ref{Thm:id_primary}.
Next, we show the equivalence of the three identification formulas.
For the second identification formula, we have 
\begin{align*}
\E\left\{ \frac{SRY}{\pi_{R_{1}}(X)}\right\}  & =\E\left\{ \frac{\pi_{S}(X)\pi_{R_{1}}(X)\mu_{1}(X)}{\pi_{R_{1}}(X)}\right\} =\E\{\pi_{S}(X)\mu_{1}(X)\},\\
\E\left\{ \frac{(1-S)Rq_{S}(X)Y}{\pi_{R_{0}}(X)}\right\}  & =\E\left\{ \frac{\{1-\pi_{S}(X)\}\pi_{R_{0}}(X)q_{S}(X)\mu_{0}(X)}{\pi_{R_{0}}(X)}\right\} =\E\{\pi_{S}(X)\mu_{0}(X)\},
\end{align*}
which holds under the assumption of EC outcome comparability in Table
\ref{tab:Key-assumptions}. For the third identification formula,
we have 
\begin{align*}
\E\left\{ S\pi_{R_{1}}(X)Y+S\{1-\pi_{R_{1}}(X)\}\mu_{1}(X)\right\}  & =\E\{\pi_{S}(X)\mu_{1}(X)\},\\
\E\left\{ S\mu_{0}(X)\right\}  & =\E\{\pi_{S}(X)\mu_{0}(X)\},
\end{align*}
which completes the proof of Theorem \ref{Thm:id_primary}.

\subsection{Proof of Theorem \ref{thm:id_tilting}}

Under Assumptions 1 and 2 in Table \ref{tab:Key-assumptions}, and
Model \ref{assump:tilting-MNAR} with fixed $\gamma_{R_{0}}$,
$\gamma_{R_{1}}$ and $\gamma_{S}$, the parameter $\tau=\E\{Y\{1,R(1)\}-Y\{0,R(0)\}\mid S=1\}$
is identifiable. For the term $\E[Y\{1,R(1)\}\mid S=1]$, we have
\begin{align*}
 & \E[Y\{1,R(1)\}\mid S=1] \\
 & =\frac{1}{P(S=1)}\E\left(\E[SR(1)Y(1,1)+S\{1-R(1)\}Y(1,0)\mid X]\right)\\
 & =\frac{1}{P(S=1)}\E\left[\pi_{S}(X)\E\{R(1)\mid X,S=1\}\E\{Y(1,1)\mid X,R(1)=1,S=1\}\right]\\
 & +\frac{1}{P(S=1)}\E\left[\pi_{S}(X)\E\{1-R(1)\mid X,S=1\}\E\{Y(1,0)\mid X,R=0,S=1\}\right],
\end{align*}
where $\E\{Y(1,1)\mid X,R(1)=1,S=1\}=\mu_{1}(X)$, and
\begin{align*}
\E\{Y(1,0)\mid X,R=0,S=1\} & =\int y{\rm d}F\{Y(1,0)\mid X,S=1,R=0\}\\
 & =\int y{\rm d}F\{Y(1,1)\mid X,S=1,R=1\}\frac{\exp\{\gamma_{R_{1}}y\}}{c(X;\gamma_{R_{1}})}\\
 & =\frac{\E[Y\exp\{\gamma_{R_{1}}Y(1)\}\mid X,S=1,R=1)]}{c(X;\gamma_{R_{1}})}=\frac{b(X;\gamma_{R_{1}})}{c(X;\gamma_{R_{1}})}.
\end{align*}
Thus, we have 
\begin{align}
\E[Y\{1,R(1)\}\mid S=1] & =\frac{1}{P(S=1)}\E\left[\pi_{S}(X)\pi_{R_{1}}(X)\mu_{1}(X)+\pi_{S}(X)\{1-\pi_{R_{1}}(X)\}\frac{b(X;\gamma_{R_{1}})}{c(X;\gamma_{R_{1}})}\right]\label{eq:ident_Y1_sens}
\end{align}
For the term $\E[Y\{0,R(0)\}\mid S=1]$, we have 
\begin{equation}
\E[Y\{0,R(0)\}\mid S=1]=\frac{1}{P(S=1)}\E[\pi_{S}(X)\E\{Y(0)\mid X,S=1\}],\label{eq:ident_Y0_sens}
\end{equation}
where 
\begin{align*}
\E\{Y(0)\mid X,S=1\} & =\int y{\rm d}F\{Y(0)\mid X,S=1\}\\
 & =\int y{\rm d}F\{Y(0)\mid X,S=0\}\frac{\exp\{\gamma_{S}Y(0)\}}{\E[\exp\{\gamma_{S}Y(0)\}\mid X,S=0]}\\
 & =\frac{\E[Y\exp\{\gamma_{S}Y(0)\}\mid X,S=0)]}{\E[\exp\{\gamma_{S}Y(0)\}\mid X,S=0]}.
\end{align*}
Further, we show that 
\begin{align*}
 & \E[\exp\{\gamma_{S}Y(0)\}\mid X,S=0] \\
 & =\E[\exp\{\gamma_{S}Y(0)\}\mid X,S=0)]\\
 & =\pi_{R_{0}}(X)\E[\exp\{\gamma_{S}Y(0)\}\mid X,S=0,R=1)]\\
 & +\{1-\pi_{R_{0}}(X)\}\int y\exp\{\gamma_{S}Y(0)\}{\rm d}F\{Y(0,0)\mid X,S=0,R=0\}\\
 & =\pi_{R_{1}}(X)c(V;\gamma_{S})\\
 & +\{1-\pi_{R_{0}}(X)\}\int y\exp\{\gamma_{S}Y(0)\}{\rm d}F\{Y(0,1)\mid X,S=0,R=1\}\frac{\exp\{\gamma_{R_{0}}Y(0)\}}{c(X;\gamma_{R_{0}})}\\
 & =\pi_{R_{1}}(X)c(V;\gamma_{S})+\{1-\pi_{R_{0}}(X)\}\frac{c(X;\gamma_{S}+\gamma_{R_{0}})}{c(X;\gamma_{R_{0}})},
\end{align*}
and 
\[
\E[Y\exp\{\gamma_{S}Y(0)\}\mid X,S=0)]=\pi_{R_{1}}(X)b(V;\gamma_{S})+\{1-\pi_{R_{0}}(X)\}\frac{b(X;\gamma_{S}+\gamma_{R_{0}})}{c(X;\gamma_{R_{0}})}.
\]
Let $d(V;\gamma_{R_{0}},\gamma_{S})$ and $e(V;\gamma_{R_{0}},\gamma_{S})$
be defined as in Table \ref{tab:Key-assumptions}(B), we have $\E\{Y(0)\mid X,S=1\}=d(V;\gamma_{R_{0}},\gamma_{S})/e(V;\gamma_{R_{0}},\gamma_{S})$.
Combining the terms (\ref{eq:ident_Y1_sens}) and (\ref{eq:ident_Y0_sens}), the parameter $\tau$ is identified under our sensitivity framework:
\begin{align*}
\tau & =\frac{1}{P(S=1)}\E\left[\pi_{S}(X)\pi_{R_{1}}(X)\mu_{1}(X)+\pi_{S}(X)\{1-\pi_{R_{1}}(X)\}\frac{b(X;\gamma_{R_{1}})}{c(X;\gamma_{R_{1}})}\right]\\
 & -\frac{1}{P(S=1)}\E\left\{ \frac{\pi_{S}(X)d(X;\gamma_{R_{0}},\gamma_{S})}{e(X;\gamma_{R_{0}},\gamma_{S})}\right\} ,
\end{align*}
which completes the proof of Theorem \ref{thm:id_tilting}.

\subsection{Proof of Theorem \ref{thm:EIF-tilting}}

\subsubsection{Preliminaries}

We will use the semi-parametric theory in \cite{tsiatis2006semiparametric}
to derive the EIFs under the tilting framework. The EIF for the primary
analysis in Theorem \ref{Thm:eif0} can be obtained by letting $\gamma_{S}=\gamma_{R_{0}}=\gamma_{R_{1}}=0$.
Let $V=(X,A,R,Y,S)$ be the random vector of all observed variables
with the likelihood factorized as
\begin{align*}
f(V) & =f(X)f(S\mid X)\{f(R\mid X,S=1)f(Y\mid X,R,S=1)\}^{S}\\
 & \times\{f(R\mid X,S=0)f(Y\mid X,R,S=0)\}^{1-S}.
\end{align*}
To derive the EIFs, we consider a one-dimensional parametric submodel
$f_{\theta}(V)$, which contains the true model $f(V)$ at $\theta=0$,
i.e., $f_{\theta}(V)\mid_{\theta=0}=f(V)$. Let $s_{\theta}(V)$ be
the score function of the submodel. From the factorization of the
likelihood, the score function $s_{\theta}(V)$ can be decomposed
as 
\begin{align*}
s_{\theta}(V) & =s_{\theta}(X)+s_{\theta}(S\mid X)+S\times s_{\theta}(R\mid X,S=1)+S\times s_{\theta}(Y\mid X,R,S=1)\\
 & +(1-S)\times s_{\theta}(R\mid X,S=0)+(1-S)\times s_{\theta}(Y\mid X,R,S=0),
\end{align*}
where $s_{\theta}(X)=\partial\log f_{\theta}(X)/\partial\theta$,
$s_{\theta}(S\mid X)=\partial\log f_{\theta}(S\mid X)/\partial\theta$,
$s_{\theta}(R\mid X,S=1)=\partial\log f_{\theta}(R\mid X,S=1)$, $s_{\theta}(Y\mid X,R,S=1)=\partial\log f_{\theta}(Y\mid X,R,S=1)$,
$s_{\theta}(R\mid X,S=0)=\partial\log f_{\theta}(R\mid X,S=0)$, $s_{\theta}(Y\mid X,R,S=0)=\partial\log f_{\theta}(Y\mid X,R,S=0)$
are the score functions corresponding to the components of the factorized
likelihood. From the semiparametric theory, the tangent space that
is the mean-square closure of the space spanned by the score vector
$s_{\theta}(V)$ under the parametric submodel is defined as
\[
\Lambda=H_{1}\oplus H_{2}\oplus H_{3}\oplus H_{4},
\]
where 
\begin{align*}
H_{1} & =\{\Gamma(X):\E\{\Gamma(X)\}=0\},\\
H_{2} & =\left\{ \{S-\pi_{S}(X)\}a(X)\right\} ,\\
H_{3} & =\left\{ S\{R-\pi_{R_{1}}(X)\}b(X)\right\} \oplus\left\{ (1-S)\{R-\pi_{R_{0}}(X)\}c(X)\right\} ,\\
H_{4} & =\left\{ S\Gamma(Y,X,R):\E\{\Gamma(Y,X,R)\mid X,R,S\}=0\right\} \\
 &\oplus\left\{ (1-S)\Gamma(Y,X,R):\E\{\Gamma(Y,X,R)\mid X,R,S\}=0\right\} .
\end{align*}
for any arbitrary square-integrable measurable functions $a(X)$,
$b(X)$, and $c(X)$. The EIF for $\tau$, denoted by $\phi_{\text{eff}}^{t}(V;P_{0})$, should satisfy $\dot{\tau}_{\theta}\mid_{\theta=0}=\E\{\phi_{\text{eff}}^{t}(V)s(V)\}$,
where $s(V)=s_{\theta}(V)\mid_{\theta=0}$ is the score function evaluated
at the true parameter $\theta=0$. 

\subsubsection{EIF under the tilting sensitivity models}

The derivation of the EIF for $\tau=\tau_{1}-\tau_{0}$ is constituted by two parts. 
\paragraph{Step 1}
The first part is to derive the EIF for $\tau_{1}=\E[Y\{1,R(1)\}\mid S=1]=\E\{N_{1,\theta}(V)\}/D_{\theta}(V)$,
where 
\begin{align*}
N_{1,\theta}(V) & =\pi_{S}(X)\pi_{R_{1}}(X)\mu_{1}(X)+\pi_{S}(X)\{1-\pi_{R_{1}}(X)\}b(X;\gamma_{R_{1}})/c(X;\gamma_{R_{1}}),\\
D_{\theta}(V) & =P(S=1),
\end{align*}
by Theorem \ref{thm:id_tilting}. From the chain rule, we can show
\begin{align*}
\dot{N}_{1,\theta}(V)|_{\theta=0} & =\frac{\partial}{\partial\theta}\E_{\theta}\left[\pi_{S}(X)\pi_{R_{1}}(X)\mu_{1}(X)+\pi_{S}(X)\{1-\pi_{R_{1}}(X)\}\frac{b(X;\gamma_{R_{1}})}{c(X;\gamma_{R_{1}})}\right]\\
 & =\E_{\theta}\{\dot{\pi}_{S}(X)\pi_{R_{1}}(X)\mu_{1}(X)+\pi_{S}(X)\dot{\pi}_{R_{1}}(X)\mu_{1}(X)\}|_{\theta=0}\\
 & +\E_{\theta}\{\pi_{S}(X)\pi_{R_{1}}(X)\dot{\mu}_{1}(X)+\pi_{S}(X)\pi_{R_{1}}(X)\mu_{1}(X)s(X)\}|_{\theta=0}\\
 & +\E_{\theta}[\dot{\pi}_{S}(X)\{1-\pi_{R_{1}}(X)\}\frac{b(X;\gamma_{R_{1}})}{c(X;\gamma_{R_{1}})}-\pi_{S}(X)\dot{\pi}_{R_{1}}(X)\frac{b(X;\gamma_{R_{1}})}{c(X;\gamma_{R_{1}})}\\
 & +\pi_{S}(X)\{1-\pi_{R_{1}}(X)\}\frac{\partial}{\partial\theta}\frac{b(X;\gamma_{R_{1}})}{c(X;\gamma_{R_{1}})}+\pi_{S}(X)\{1-\pi_{R_{1}}(X)\}\frac{b(X;\gamma_{R_{1}})}{c(X;\gamma_{R_{1}})}s(X)]|_{\theta=0},
\end{align*}
where $\E_{\theta}\{\dot{\pi}_{S}(X)\}\mid_{\theta=0}=\E_{\theta}[\{S-\pi_{S}(X)\}s_{\theta}(S\mid X)]\mid_{\theta=0}$,
\begin{align*}
\E_{\theta}\{\dot{\pi}_{R_{1}}(X)\}\mid_{\theta=0} & =\E[\{R-\pi_{R_{1}}(X)\}s(R\mid X,S=1)\mid X,S=1]\\
 & =\E\left[\frac{S\{R-\pi_{R_{1}}(X)\}}{\pi_{S}(X)}s(R\mid X,S)\mid X\right],
\end{align*}
\begin{align*}
\E_{\theta}\{\dot{\mu}_{1}(X)\}\mid_{\theta=0} & =\E[\{Y-\mu_{1}(X)\}s(Y\mid X,S=1,R=1)\mid X,S=1,R=1]\\
 & =\E\left[\frac{SR\{Y-\mu_{1}(X)\}}{\pi_{S}(X)\pi_{R_{1}}(X)}s(Y\mid X,S,R)\mid X\right],
\end{align*}
$b(X;\gamma_{R_{1}})=\E[Y\exp\{\gamma_{R_{1}}Y(1)\}\mid X,S=1,R=1]$, and
\begin{align*}
 & \E_{\theta}\left\{ \frac{\partial}{\partial\theta}\frac{b(X;\gamma_{R_{1}})}{c(X;\gamma_{R_{1}})}\right\} \mid_{\theta=0}\\
 & =\E\left[\frac{Y\exp\{\gamma_{R_{1}}Y(1)\}s(Y\mid X,S,R)}{c(X;\gamma_{R_{1}})}\mid X,S=1,R=1]\right]\\
 & -\E\left[\frac{\exp\{\gamma_{R_{1}}Y(1)\}s(Y\mid X,S,R)b(X;\gamma_{R_{1}})}{\{c(X;\gamma_{R_{1}})\}^{2}}\mid X,S=1,R=1\right]\\
 & =\E\left[\frac{SRY\exp\{\gamma_{R_{1}}Y(1)\}c(X;\gamma_{R_{1}})-SRb(X;\gamma_{R_{1}})\exp\{\gamma_{R_{1}}Y(1)\}}{\pi_{S}(X)\pi_{R_{1}}(X)\{c(X;\gamma_{R_{1}})\}^{2}}s(Y\mid X,S,R)\mid X\right].
\end{align*}
Collecting these terms together, the EIF of $N_{\theta}$ is 
\begin{align*}
\psi_{N_{1,\theta}}(V) & =SRY+S(1-R)\frac{b(X;\gamma_{R_{1}})}{c(X;\gamma_{R_{1}})}\\
& +SRq_{R}(1,X)\left[\frac{Y\exp\{\gamma_{R_{1}}Y(1)\}}{c(X;\gamma_{R_{1}})}-\frac{b(X;\gamma_{R_{1}})\exp\{\gamma_{R_{1}}Y(1)\}}{\{c(X;\gamma_{R_{1}})\}^{2}}\right],
\end{align*}
where $q_{R_{s}}(X)=\{1-\pi_{R_{1}}(X)\}/\pi_{R_{1}}(X)$. For the denominator $D_{\theta}(V)$, we have $\partial D_{\theta}(V)/\partial\theta\mid_{\theta=0}=\E\{S\partial P_{\theta}(S=1)/\partial\theta\}\mid_{\theta=0}$.
Therefore, the EIF for the first part $\tau_{1}=\E[Y\{1,R(1)\}\mid S=1]$ is 
\begin{align}
\psi_{\tau_{1}}^{t}(V;P_{0}) & =\frac{\psi_{N_{1,\theta}}(V)-\tau_{1}S}{P(S=1)}\nonumber \\
 & =\frac{S}{P(S=1)}RY+\frac{S}{P(S=1)}\left[(1-R)\frac{b(X;\gamma_{R_{1}})}{c(X;\gamma_{R_{1}})}+Rq_{R_{1}}(X)g(V;\gamma_{R_{1}})\right]\label{eq:EIF_tau1_tilting} \\
 &-\frac{S \tau_{0}}{P(S=1)},\nonumber
\end{align}
where 
\[
g(V;\gamma_{R_{1}})=\frac{Y\exp\{\gamma_{R_{1}}Y(1)\}}{c(X;\gamma_{R_{1}})}-\frac{b(X;\gamma_{R_{1}})\exp\{\gamma_{R_{1}}Y(1)\}}{\{c(X;\gamma_{R_{1}})\}^{2}},
\]
and $\E\{g(V;\gamma_{R_{1}})\}=0$.

\paragraph{part 2}
The second part is to derive the
EIF of $\tau_{0}=\E[Y\{0,R(0)\}\mid S=1]=\E\{N_{0,\theta}(V)\}/D_{\theta}(V)$.
We can show the pathwise derivative of $N_{0,\theta}(V)$ is 
\begin{align*}
\dot{N}_{0,\theta}(V) & =\frac{\partial}{\partial\theta}\E_{\theta}\left\{ \frac{\pi_{S}(X)d(X;\gamma_{R_{0}},\gamma_{S})}{e(X;\gamma_{R_{0}},\gamma_{S})}\right\} \mid_{\theta=0}\\
 & =\E_{\theta}\left\{ \dot{\pi}_{S}(X)\frac{d(X;\gamma_{R_{0}},\gamma_{S})}{e(X;\gamma_{R_{0}},\gamma_{S})}+\pi_{S}(X)\frac{\partial}{\partial\theta}\frac{d(X;\gamma_{R_{0}},\gamma_{S})}{e(X;\gamma_{R_{0}},\gamma_{S})}+\frac{\pi_{S}(X)d(X;\gamma_{R_{0}},\gamma_{S})}{e(X;\gamma_{R_{0}},\gamma_{S})}\right\} .
\end{align*}
We can show that 
\begin{align*}
\E_{\theta}\left\{ \frac{\partial}{\partial\theta}\frac{d(X;\gamma_{R_{0}},\gamma_{S})}{e(X;\gamma_{R_{0}},\gamma_{S})}\right\} \mid_{\theta=0} & =\E_{\theta}\left[\frac{\partial d(X;\gamma_{R_{0}},\gamma_{S})/\partial\theta}{e(X;\gamma_{R_{0}},\gamma_{S})}-\frac{\partial e(X;\gamma_{R_{0}},\gamma_{S})/\partial\theta}{\{e(X;\gamma_{R_{0}},\gamma_{S})\}^{2}}d(X;\gamma_{R_{0}},\gamma_{S})\right]\mid_{\theta=0}.
\end{align*}
Let 
\begin{align*}
m_{1}(V;\gamma_{S},\gamma_{R_{0}}) & =b(V;\gamma_{S})c(V;\gamma_{R_{0}})-b(V;\gamma_{S}+\gamma_{R_{0}})\\
m_{2}(V;\gamma_{S},\gamma_{R_{0}}) & =c(V;\gamma_{S})c(V;\gamma_{R_{0}})-c(V;\gamma_{S}+\gamma_{R_{0}})\\
m_{3}(V;\gamma_{S},\gamma_{R_{0}}) & =b(V;\gamma_{S})\exp\{\gamma_{R_{0}}Y(0)\}+c(V;\gamma_{R_{0}})[Y\exp\{\gamma_{S}Y(0)\}-b(V;\gamma_{S})]\\
 & +q_{R_{0}}(X)Y\exp\{(\gamma_{S}+\gamma_{R_{0}})Y(0)\}\\
m_{4}(V;\gamma_{S},\gamma_{R_{0}}) & =c(V;\gamma_{S})\exp\{\gamma_{R_{0}}Y(0)\}+c(V;\gamma_{R_{0}})\{\exp[\gamma_{S}Y(0)]-c(V;\gamma_{S})\}\\
 & +q_{R_{0}}(X)\exp\{(\gamma_{S}+\gamma_{R_{0}})Y(0)\},
\end{align*}
we can show that 
\begin{align*}
\E_{\theta}\left\{ \frac{\partial d(V;\gamma_{R_{0}},\gamma_{S})}{\partial\theta}\right\} \mid_{\theta=0} & =\E_{\theta}\left[\left\{ \frac{(1-S)\{R-\pi_{R_{0}}(X)\}m_{1}(V;\gamma_{S},\gamma_{R_{0}})}{1-\pi_{S}(X)}\right\} s_{\theta}(V)\right]\\
 & +\E_{\theta}\left[\left\{ \frac{(1-S)Rm_{3}(V;\gamma_{S},\gamma_{R_{0}})}{1-\pi_{S}(X)}\right\} s_{\theta}(V)\right],
\end{align*}
and
\begin{align*}
\E_{\theta}\left\{ \frac{\partial e(V;\gamma_{R_{0}},\gamma_{S})}{\partial\theta}\right\} \mid_{\theta=0} & =\E_{\theta}\left[\left\{ \frac{(1-S)\{R-\pi_{R_{0}}(X)\}m_{2}(V;\gamma_{S},\gamma_{R_{0}})}{1-\pi_{S}(X)}\right\} s_{\theta}(V)\right]\\
 & +\E_{\theta}\left[\left\{ \frac{(1-S)Rm_{4}(V;\gamma_{S},\gamma_{R_{0}})}{1-\pi_{S}(X)}\right\} s_{\theta}(V)\right].
\end{align*}
Therefore, the EIF for the second part $\tau_{0}=\E[Y\{0,R(0)\}\mid S=1]$
is 
\begin{align}
\psi_{\tau_{0}}^{t}(V;P_{0}) & =\frac{\psi_{N_{0,\theta}}(V)-\tau_{0}S}{P(S=1)}\nonumber \\
 & =\frac{S}{P(S=1)}\frac{d(X;\gamma_{R_{0}},\gamma_{S})}{e(X;\gamma_{R_{0}},\gamma_{S})}+\frac{1-S}{P(S=1)}q_{S}(X)h(V;\gamma_{R_{0}},\gamma_{S})-\frac{S\tau_{0}}{P(S=1)},\label{eq:EIF_tau0_tilting}
\end{align}
where 
\begin{align*}
h(V;\gamma_{R_{0}},\gamma_{S}) & =\frac{\{R-\pi_{R_{0}}(X)\}m_{1}(V;\gamma_{S},\gamma_{R_{0}})}{e(V;\gamma_{R_{0}},\gamma_{S})}\\
 & -\frac{\{R-\pi_{R_{0}}(X)\}d(V;\gamma_{R_{0}},\gamma_{S})m_{2}(V;\gamma_{S},\gamma_{R_{0}})}{e(V;\gamma_{R_{0}},\gamma_{S})^{2}}\\
 & +\frac{Rm_{3}(V;\gamma_{S},\gamma_{R_{0}})}{e(V;\gamma_{R_{0}},\gamma_{S})}-\frac{Rd(V;\gamma_{R_{0}},\gamma_{S})m_{4}(V;\gamma_{S},\gamma_{R_{0}})}{e(V;\gamma_{R_{0}},\gamma_{S})^{2}}.
\end{align*}
Combining (\ref{eq:EIF_tau0_tilting}) and (\ref{eq:EIF_tau1_tilting}),
it gives us the EIF for $\tau$ as 
\begin{align*}
\phi_{\text{eff}}^{t}(V;P_{0}) & =\psi_{\tau_{1}}^{t}(V;P_{0})-\psi_{\tau_{0}}^{t}(V;P_{0})=\frac{S}{P(S=1)}RY\\
 & +\frac{S}{P(S=1)}\left[(1-R)\frac{b(X;\gamma_{R_{1}})}{c(X;\gamma_{R_{1}})}+Rq_{R_{1}}(X)g(V;\gamma_{R_{1}})\right]\\
 & -\frac{1}{P(S=1)}\left\{ S\frac{d(X;\gamma_{R_{0}},\gamma_{S})}{e(X;\gamma_{R_{0}},\gamma_{S})}+(1-S)q_{S}(X)h(V;\gamma_{R_{0}},\gamma_{S})\right\} -\frac{S\tau}{P(S=1)},
\end{align*}
which completes the proof of Theorem \ref{thm:EIF-tilting}. 

\subsection{Proof of Theorem \ref{thm:EIF-tilting-1}}

\begin{assumption}\label{assump:regularity}
\begin{enumerate}
\item [(C1)] $\|\widehat{\pi}_{S}(X)-\pi_{S}(X)\|_{L_{2}}=o_{\pr}(1)$, $\|\widehat{\pi}_{R_{s}}(X)-\pi_{R_{s}}(X)\|_{L_{2}}=o_{\pr}(1)$,
$\|\widehat{b}(X;\gamma)-b(X;\gamma)\|_{L_{2}}=o_{\pr}(1)$, and $\|\widehat{c}(X;\gamma)-c(X;\gamma)\|_{L_{2}}=o_{\pr}(1)$
for any $\gamma$;
\item [(C2)] $\phi_{\text{eff}}^{t}(V)$ belongs to a Donsker class \citep{wellner2013weak};
\item [(C3)] $0<c_{1}\leq\pi_{S}(X),\pi_{R_{a}}(X)\leq c_{2}<1$ and their
estimated counterparts are bounded away from $0$ and $1$ for some
constants $c_{1}$ and $c_{2}.$
\end{enumerate}
\end{assumption}To investigate the asymptotic properties of $\widehat{\tau}^{t}$,
we show that $\widehat{\tau}^{t}-\tau$ is constituted by three components
\begin{align}
\widehat{\tau}^{t}-\tau & =\int\phi_{\text{eff}}^{t}(V;\widehat{P})d\mathbb{P}_{N}-\int\phi_{\text{eff}}^{t}(V;\widehat{P})d\mathbb{P}\nonumber \\
 & =\int\phi_{\text{eff}}^{t}(V;P_{0})d\mathbb{P}_{N}\label{eq:asym_part1}\\
 & +\int\{\phi_{\text{eff}}^{t}(V;\widehat{\mathbb{P}})-\phi_{\text{eff}}^{t}(V;P_{0})\}d\mathbb{P}\label{eq:eq:asym_part2}\\
 & +\int\{\phi_{\text{eff}}^{t}(V;\widehat{\mathbb{P}})-\phi_{\text{eff}}^{t}(V;P_{0})\}d(\mathbb{P}_{N}-\mathbb{P}),\label{eq:asym_part3}
\end{align}
where $\mathbb{P}_{N}$ is the empirical counterpart of the true distribution
$\mathbb{P}$. By the central limit theorem, the first part (\ref{eq:asym_part1})
is asymptotically normal. The third part (\ref{eq:asym_part3}) is
the empirical process which is ensured to be negligible under Assumption
(C2). Thus, if the second part (\ref{eq:eq:asym_part2}) is negligible
(i.e., $o_{\pr}(N^{-1/2})$), the asymptotic properties of $\widehat{\tau}^{t}-\tau$
can then be established. The second part is the second-order remainder
error, which can be characterized by: 
\begin{align}
 & \int\{\phi_{\text{eff}}^{t}(V;\widehat{\mathbb{P}})-\phi_{\text{eff}}^{t}(V;P_{0})\}d\mathbb{P}\nonumber \\
 & =\mathbb{P}\left[\pi_{S}(X)\{1-\pi_{R_{1}}(X)\}\frac{\widehat{b}(X;\gamma_{R_{1}})}{\widehat{c}(X;\gamma_{R_{1}})}+\pi_{S}(X)\pi_{R_{1}}(X)\widehat{q}_{R_{1}}(X)\widehat{g}(V;\gamma_{R_{1}})\right]\nonumber \\
 & -\mathbb{P}\left[\pi_{S}(X)\frac{\widehat{d}(X;\gamma_{R_{0}},\gamma_{S})}{\widehat{e}(X;\gamma_{R_{0}},\gamma_{S})}+\{1-\pi_{S}(X)\}\widehat{q}_{S}(X)\widehat{h}(V;\gamma_{R_{0}},\gamma_{S})\right]\nonumber \\
 & -\mathbb{P}\left[\pi_{S}(X)\{1-\pi_{R_{1}}(X)\}\frac{b(X;\gamma_{R_{1}})}{c(X;\gamma_{R_{1}})}\right]+\mathbb{P}\left\{ \pi_{S}(X)\frac{d(X;\gamma_{R_{0}},\gamma_{S})}{e(X;\gamma_{R_{0}},\gamma_{S})}\right\} \nonumber \\
 & =\mathbb{P}\left(\pi_{S}(X)\pi_{R_{1}}(X)\left[\widehat{q}_{R_{1}}(X)\widehat{g}(V;\gamma_{R_{1}})-q_{R_{1}}(X)\left\{ \frac{b(X;\gamma_{R_{1}})}{c(X;\gamma_{R_{1}})}-\frac{\widehat{b}(X;\gamma_{R_{1}})}{\widehat{c}(X;\gamma_{R_{1}})}\right\} \right]\right)\label{eq:remainder-part1}\\
 & +\mathbb{P}\left(\{1-\pi_{S}(X)\}\left[q_{S}(X)\left\{ \frac{d(X;\gamma_{R_{0}},\gamma_{S})}{e(X;\gamma_{R_{0}},\gamma_{S})}-\frac{\widehat{d}(X;\gamma_{R_{0}},\gamma_{S})}{\widehat{e}(X;\gamma_{R_{0}},\gamma_{S})}\right\} +\widehat{q}_{S}(X)\widehat{h}(V;\gamma_{R_{0}},\gamma_{S})\right]\right).\label{eq:remainder-part2}
\end{align}
The first part (\ref{eq:remainder-part1}) can be written as 
\begin{align*}
 & \left|\mathbb{P}\left(\pi_{S}(X)\pi_{R_{1}}(X)\left[\widehat{q}_{R_{1}}(X)\widehat{g}(V;\gamma_{R_{1}})-q_{R_{1}}(X)\left\{ \frac{b(X;\gamma_{R_{1}})}{c(X;\gamma_{R_{1}})}-\frac{\widehat{b}(X;\gamma_{R_{1}})}{\widehat{c}(X;\gamma_{R_{1}})}\right\} \right]\right)\right|\\
= & \left|\mathbb{P}\left[\frac{b(X;\gamma_{R_{1}})\widehat{c}(X;\gamma_{R_{1}})-\widehat{b}(X;\gamma_{R_{1}})c(X;\gamma_{R_{1}})}{\widehat{c}(X;\gamma_{R_{1}})}\left\{ \frac{\widehat{q}_{R_{1}}(X)}{\widehat{c}(X;\gamma_{R_{1}})}-\frac{q_{R_{1}}(X)}{c(X;\gamma_{R_{1}})}\right\} \right]\right|\\
\leq & C_{1}\|\widehat{c}(X;\gamma_{R_{1}})-c(X;\gamma_{R_{1}})\|_{L_{2}}^{2}\\
 & +C_{2}\|\widehat{c}(X;\gamma_{R_{1}})-c(X;\gamma_{R_{1}})\|_{L_{2}}\times\|\widehat{b}(X;\gamma_{R_{1}})-b(X;\gamma_{R_{1}})\|_{L_{2}}\\
 & +C_{3}\|\widehat{q}_{R_{1}}(X)-q_{R_{1}}(X)\|_{L_{2}}\times\|\widehat{c}(X;\gamma_{R_{1}})-c(X;\gamma_{R_{1}})\|_{L_{2}}\\
 & +C_{4}\|\widehat{q}_{R_{1}}(X)-q_{R_{1}}(X)\|_{L_{2}}\times\|\widehat{b}(X;\gamma_{R_{1}})-b(X;\gamma_{R_{1}})\|_{L_{2}}\\
\lesssim & \left\{ \|\widehat{q}_{R_{1}}(X)-q_{R_{1}}(X)\|_{L_{2}}+\|\widehat{c}(X;\gamma_{R_{1}})-c(X;\gamma_{R_{1}})\|_{L_{2}}\right\} \\
 & \times\left\{ \|\widehat{c}(X;\gamma_{R_{1}})-c(X;\gamma_{R_{1}})\|_{L_{2}}+\|\widehat{b}(X;\gamma_{R_{1}})-b(X;\gamma_{R_{1}})\|_{L_{2}}\right\},
\end{align*}
where
\begin{align*}
\widehat{g}(V;\gamma_{R_{1}}) & =\frac{b(X;\gamma_{R_{1}})}{\widehat{c}(X;\gamma_{R_{1}})}-\frac{\widehat{b}(X;\gamma_{R_{1}})c(X;\gamma_{R_{1}})}{\{\widehat{c}(X;\gamma_{R_{1}})\}^{2}}.
\end{align*}
Let $\widehat{m}_{1}(V;\gamma_{S},\gamma_{R_{0}})$, $\widehat{m}_{2}(V;\gamma_{S},\gamma_{R_{0}})$,
$\widehat{m}_{3}(V;\gamma_{S},\gamma_{R_{0}})$, and $\widehat{m}_{4}(V;\gamma_{S},\gamma_{R_{0}})$
be their estimated counterparts, we can show
\begin{align*}
\widehat{h}(V;\gamma_{R_{0}},\gamma_{S}) & =\frac{\{\pi_{R_{0}}(X)-\widehat{\pi}_{R_{0}}(X)\}\widehat{m}_{1}(V;\gamma_{S},\gamma_{R_{0}})}{\widehat{e}(V;\gamma_{R_{0}},\gamma_{S})}\\
 & -\frac{\{\pi_{R_{0}}(X)-\widehat{\pi}_{R_{0}}(X)\}\widehat{d}(V;\gamma_{R_{0}},\gamma_{S})\widehat{m}_{2}(V;\gamma_{S},\gamma_{R_{0}})}{\widehat{e}(V;\gamma_{R_{0}},\gamma_{S})^{2}}\\
 & +\frac{\pi_{R_{0}}(X)\widehat{m}_{3}(V;\gamma_{S},\gamma_{R_{0}})}{\widehat{e}(V;\gamma_{R_{0}},\gamma_{S})}-\frac{\pi_{R_{0}}(X)\widehat{d}(V;\gamma_{R_{0}},\gamma_{S})\widehat{m}_{4}(V;\gamma_{S},\gamma_{R_{0}})}{\widehat{e}(V;\gamma_{R_{0}},\gamma_{S})^{2}}\\
 & =\frac{\{\pi_{R_{0}}(X)-\widehat{\pi}_{R_{0}}(X)\}\{\widehat{e}(V;\gamma_{R_{0}},\gamma_{S})\widehat{m}_{1}(V;\gamma_{S},\gamma_{R_{0}})-\widehat{d}(V;\gamma_{R_{0}},\gamma_{S})\widehat{m}_{2}(V;\gamma_{S},\gamma_{R_{0}})\}}{\widehat{e}(V;\gamma_{R_{0}},\gamma_{S})^{2}}\\
 & +\frac{d(V;\gamma_{R_{0}},\gamma_{S})\widehat{e}(V;\gamma_{R_{0}},\gamma_{S})-e(V;\gamma_{R_{0}},\gamma_{S})\widehat{d}(V;\gamma_{R_{0}},\gamma_{S})}{\widehat{e}(V;\gamma_{R_{0}},\gamma_{S})^{2}}\\
 & +\frac{\Delta_{m_{3}}\widehat{e}(V;\gamma_{R_{0}},\gamma_{S})-\Delta_{m_{4}}\widehat{d}(V;\gamma_{R_{0}},\gamma_{S})}{\widehat{e}(V;\gamma_{R_{0}},\gamma_{S})^{2}},
\end{align*}
where
\begin{align*}
\Delta_{m_{3}}= & \pi_{R_{0}}(X)\{\widehat{b}(V;\gamma_{S})-b(V;\gamma_{S})\}\{c(V;\gamma_{R_{0}})-\widehat{c}(V;\gamma_{R_{0}})\}\\
 & +\pi_{R_{0}}(X)\{\widehat{q}_{R_{0}}(X)-q_{R_{0}}(X)\}\{b(V;\gamma_{S}+\gamma_{R_{0}})-\widehat{b}(V;\gamma_{S}+\gamma_{R_{0}})\}\\
 & +\pi_{R_{0}}(X)\{\widehat{q}_{R_{0}}(X)-q_{R_{0}}(X)\}\widehat{b}(V;\gamma_{S}+\gamma_{R_{0}}),\\
\Delta_{m_{4}}= & \pi_{R_{0}}(X)\{\widehat{c}(V;\gamma_{S})-c(V;\gamma_{S})\}\{c(V;\gamma_{R_{0}})-\widehat{c}(V;\gamma_{R_{0}})\}\\
 & +\pi_{R_{0}}(X)\{\widehat{q}_{R_{0}}(X)-q_{R_{0}}(X)\}\{c(V;\gamma_{S}+\gamma_{R_{0}})-\widehat{c}(V;\gamma_{S}+\gamma_{R_{0}})\}\\
 & +\pi_{R_{0}}(X)\{\widehat{q}_{R_{0}}(X)-q_{R_{0}}(X)\}\widehat{c}(V;\gamma_{S}+\gamma_{R_{0}}),
\end{align*}
\begin{align*}
 & \widehat{e}(V;\gamma_{R_{0}},\gamma_{S})\widehat{m}_{1}(V;\gamma_{S},\gamma_{R_{0}})-\widehat{d}(V;\gamma_{R_{0}},\gamma_{S})\widehat{m}_{2}(V;\gamma_{S},\gamma_{R_{0}})\\
= & \widehat{c}(V;\gamma_{R_{0}})\{\widehat{c}(V;\gamma_{S}+\gamma_{R_{0}})\widehat{b}(V;\gamma_{S})-\widehat{b}(V;\gamma_{S}+\gamma_{R_{0}})\widehat{c}(V;\gamma_{S})\},
\end{align*}
\begin{align*}
 & \pi_{R_{0}}(X)\{\widehat{q}_{R_{0}}(X)-q_{R_{0}}(X)\}\widehat{b}(V;\gamma_{S}+\gamma_{R_{0}})\widehat{e}(X;\gamma_{R_{0}},\gamma_{S})\\
- & \pi_{R_{0}}(X)\{\widehat{q}_{R_{0}}(X)-q_{R_{0}}(X)\}\widehat{c}(V;\gamma_{S}+\gamma_{R_{0}})\widehat{d}(X;\gamma_{R_{0}},\gamma_{S})\\
= & \pi_{R_{0}}(X)\{\widehat{q}_{R_{0}}(X)-q_{R_{0}}(X)\}\widehat{\pi}_{R_{0}}(X)\widehat{c}(V;\gamma_{R_{0}})\\
 & \times\{\widehat{b}(V;\gamma_{S}+\gamma_{R_{0}})\widehat{c}(V;\gamma_{S})-\widehat{c}(V;\gamma_{S}+\gamma_{R_{0}})\widehat{b}(V;\gamma_{S})\},
\end{align*}
and $\pi_{R_{0}}(X)\{\widehat{q}_{R_{0}}(X)-q_{R_{0}}(X)\}\widehat{\pi}_{R_{0}}(X)=\pi_{R_{0}}(X)-\widehat{\pi}_{R_{0}}(X)$.
Collecting these terms, we have
\begin{align*}
\widehat{h}(V;\gamma_{R_{0}},\gamma_{S}) & =\frac{d(V;\gamma_{R_{0}},\gamma_{S})\widehat{e}(V;\gamma_{R_{0}},\gamma_{S})-e(V;\gamma_{R_{0}},\gamma_{S})\widehat{d}(V;\gamma_{R_{0}},\gamma_{S})}{\widehat{e}(V;\gamma_{R_{0}},\gamma_{S})^{2}}\\
 & +\frac{\pi_{R_{0}}(X)\{\widehat{b}(V;\gamma_{S})-b(V;\gamma_{S})\}\{c(V;\gamma_{R_{0}})-\widehat{c}(V;\gamma_{R_{0}})\}\widehat{e}(V;\gamma_{R_{0}},\gamma_{S})}{\widehat{e}(V;\gamma_{R_{0}},\gamma_{S})^{2}}\\
 & +\frac{\pi_{R_{0}}(X)\{\widehat{q}_{R_{0}}(X)-q_{R_{0}}(X)\}\{b(V;\gamma_{S}+\gamma_{R_{0}})-\widehat{b}(V;\gamma_{S}+\gamma_{R_{0}})\}\widehat{e}(V;\gamma_{R_{0}},\gamma_{S})}{\widehat{e}(V;\gamma_{R_{0}},\gamma_{S})^{2}}\\
 & -\frac{\pi_{R_{0}}(X)\{\widehat{c}(V;\gamma_{S})-c(V;\gamma_{S})\}\{c(V;\gamma_{R_{0}})-\widehat{c}(V;\gamma_{R_{0}})\}\widehat{d}(V;\gamma_{R_{0}},\gamma_{S})}{\widehat{e}(V;\gamma_{R_{0}},\gamma_{S})^{2}}\\
 & -\frac{\pi_{R_{0}}(X)\{\widehat{q}_{R_{0}}(X)-q_{R_{0}}(X)\}\{c(V;\gamma_{S}+\gamma_{R_{0}})-\widehat{c}(V;\gamma_{S}+\gamma_{R_{0}})\}\widehat{d}(V;\gamma_{R_{0}},\gamma_{S})}{\widehat{e}(V;\gamma_{R_{0}},\gamma_{S})^{2}},
\end{align*}
and
\begin{align*}
 & q_{S}(X)\left\{ \frac{d(X;\gamma_{R_{0}},\gamma_{S})}{e(X;\gamma_{R_{0}},\gamma_{S})}-\frac{\widehat{d}(X;\gamma_{R_{0}},\gamma_{S})}{\widehat{e}(X;\gamma_{R_{0}},\gamma_{S})}\right\} +\widehat{q}_{S}(X)\widehat{h}(V;\gamma_{R_{0}},\gamma_{S})\\
= & \left\{ \frac{q_{S}(X)}{e(X;\gamma_{R_{0}},\gamma_{S})}-\frac{\widehat{q}_{S}(X)}{\widehat{e}(V;\gamma_{R_{0}},\gamma_{S})}\right\} \left\{ \frac{d(V;\gamma_{R_{0}},\gamma_{S})\widehat{e}(V;\gamma_{R_{0}},\gamma_{S})-e(V;\gamma_{R_{0}},\gamma_{S})\widehat{d}(V;\gamma_{R_{0}},\gamma_{S})}{\widehat{e}(X;\gamma_{R_{0}},\gamma_{S})}\right\} \\
 & +\widehat{q}_{S}(X)\frac{\pi_{R_{0}}(X)\{\widehat{b}(V;\gamma_{S})-b(V;\gamma_{S})\}\{c(V;\gamma_{R_{0}})-\widehat{c}(V;\gamma_{R_{0}})\}\widehat{e}(V;\gamma_{R_{0}},\gamma_{S})}{\widehat{e}(V;\gamma_{R_{0}},\gamma_{S})^{2}}\\
 & +\widehat{q}_{S}(X)\frac{\pi_{R_{0}}(X)\{\widehat{q}_{R_{0}}(X)-q_{R_{0}}(X)\}\{b(V;\gamma_{S}+\gamma_{R_{0}})-\widehat{b}(V;\gamma_{S}+\gamma_{R_{0}})\}\widehat{e}(V;\gamma_{R_{0}},\gamma_{S})}{\widehat{e}(V;\gamma_{R_{0}},\gamma_{S})^{2}}\\
 & -\widehat{q}_{S}(X)\frac{\pi_{R_{0}}(X)\{\widehat{c}(V;\gamma_{S})-c(V;\gamma_{S})\}\{c(V;\gamma_{R_{0}})-\widehat{c}(V;\gamma_{R_{0}})\}\widehat{d}(V;\gamma_{R_{0}},\gamma_{S})}{\widehat{e}(V;\gamma_{R_{0}},\gamma_{S})^{2}}\\
 & -\widehat{q}_{S}(X)\frac{\pi_{R_{0}}(X)\{\widehat{q}_{R_{0}}(X)-q_{R_{0}}(X)\}\{c(V;\gamma_{S}+\gamma_{R_{0}})-\widehat{c}(V;\gamma_{S}+\gamma_{R_{0}})\}\widehat{d}(V;\gamma_{R_{0}},\gamma_{S})}{\widehat{e}(V;\gamma_{R_{0}},\gamma_{S})^{2}}.
\end{align*}
Thus, the second part (\ref{eq:remainder-part2}) can be bounded by
\begin{align*}
 & C_{5}\|\widehat{q}_{S}(X)-q_{S}(X)\|_{L_{2}}\times\|d(V;\gamma_{R_{0}},\gamma_{S})\widehat{e}(V;\gamma_{R_{0}},\gamma_{S})-e(V;\gamma_{R_{0}},\gamma_{S})\widehat{d}(V;\gamma_{R_{0}},\gamma_{S})\|_{L_{2}}\\
 & +C_{6}\|\widehat{e}(X;\gamma_{R_{0}},\gamma_{S})-e(X;\gamma_{R_{0}},\gamma_{S})\|_{L_{2}}\times\|d(V;\gamma_{R_{0}},\gamma_{S})\widehat{e}(V;\gamma_{R_{0}},\gamma_{S})-e(V;\gamma_{R_{0}},\gamma_{S})\widehat{d}(V;\gamma_{R_{0}},\gamma_{S})\|_{L_{2}}\\
 & +C_{7}\|\widehat{b}(V;\gamma_{S})-b(V;\gamma_{S})\|_{L_{2}}\times\|c(V;\gamma_{R_{0}})-\widehat{c}(V;\gamma_{R_{0}})\|_{L_{2}}\\
 & +C_{8}\|\widehat{q}_{R_{0}}(X)-q_{R_{0}}(X)\|_{L_{2}}\times\|b(V;\gamma_{S}+\gamma_{R_{0}})-\widehat{b}(V;\gamma_{S}+\gamma_{R_{0}})\|_{L_{2}}\\
 & +C_{9}\|\widehat{c}(V;\gamma_{S})-c(V;\gamma_{S})\|_{L_{2}}\times\|c(V;\gamma_{R_{0}})-\widehat{c}(V;\gamma_{R_{0}})\|_{L_{2}}\\
 & +C_{10}\|\widehat{q}_{R_{0}}(X)-q_{R_{0}}(X)\|_{L_{2}}\times\|c(V;\gamma_{S}+\gamma_{R_{0}})-\widehat{c}(V;\gamma_{S}+\gamma_{R_{0}})\|_{L_{2}}\\
\lesssim & \sum_{\gamma\in\{\gamma_{R_{0}},\gamma_{S},\gamma_{S}+\gamma_{R_{0}}\}}\|\widehat{b}(V;\gamma)-b(V;\gamma)\|_{L_{2}}\\
 & \times\left\{ \|\widehat{q}_{R_{0}}(X)-q_{R_{0}}(X)\|_{L_{2}}+\|\widehat{q}_{S}(X)-q_{S}(X)\|_{L_{2}}+\sum_{\gamma\in\{\gamma_{R_{0}},\gamma_{S},\gamma_{S}+\gamma_{R_{0}}\}}\|\widehat{c}(V;\gamma)-c(V;\gamma)\|_{L_{2}}\right\} \\
 & +\left\{ \|\widehat{q}_{S}(X)-q_{S}(X)\|_{L_{2}}+\sum_{\gamma\in\{\gamma_{R_{0}},\gamma_{S},\gamma_{S}+\gamma_{R_{0}}\}}\|\widehat{c}(V;\gamma)-c(V;\gamma)\|_{L_{2}}\right\} \\
 & \times\left\{ \sum_{\gamma\in\{\gamma_{R_{0}},\gamma_{S},\gamma_{S}+\gamma_{R_{0}}\}}\|\widehat{c}(V;\gamma)-c(V;\gamma)\|_{L_{2}}+\sum_{\gamma\in\{\gamma_{R_{0}},\gamma_{S},\gamma_{S}+\gamma_{R_{0}}\}}\|\widehat{b}(V;\gamma)-b(V;\gamma)\|_{L_{2}}\right\} 
\end{align*}
for some constants $C_{5},\cdots,C_{10}$, where $\lesssim$ holds
up to some multiplicative constants. Putting (\ref{eq:remainder-part1})
and (\ref{eq:remainder-part2}) together, the second-order remainder of $\widehat{\tau}^{[0]}-\tau$
can be finally bounded by
\begin{align*}
 & \left[\int\{\phi_{\text{eff}}^{t}(V;\widehat{\mathbb{P}})-\phi_{\text{eff}}^{t}(V;P_{0})\}^{2}d\mathbb{P}\right]^{1/2}\\
\lesssim & \left\{ \|\widehat{q}_{R_{1}}(X)-q_{R_{1}}(X)\|_{L_{2}}+\|\widehat{c}(X;\gamma_{R_{1}})-c(X;\gamma_{R_{1}})\|_{L_{2}}\right\} \\
 & \times\left\{ \|\widehat{c}(X;\gamma_{R_{1}})-c(X;\gamma_{R_{1}})\|_{L_{2}}+\|\widehat{b}(X;\gamma_{R_{1}})-b(X;\gamma_{R_{1}})\|_{L_{2}}\right\} \\
 & +\sum_{\gamma\in\{\gamma_{R_{0}},\gamma_{S},\gamma_{S}+\gamma_{R_{0}}\}}\|\widehat{b}(V;\gamma)-b(V;\gamma)\|_{L_{2}}\\
 & \times\left\{ \|\widehat{q}_{R_{0}}(X)-q_{R_{0}}(X)\|_{L_{2}}+\|\widehat{q}_{S}(X)-q_{S}(X)\|_{L_{2}}+\sum_{\gamma\in\{\gamma_{R_{0}},\gamma_{S},\gamma_{S}+\gamma_{R_{0}}\}}\|\widehat{c}(V;\gamma)-c(V;\gamma)\|_{L_{2}}\right\} \\
 & +\left\{ \|\widehat{q}_{S}(X)-q_{S}(X)\|_{L_{2}}+\sum_{\gamma\in\{\gamma_{R_{0}},\gamma_{S},\gamma_{S}+\gamma_{R_{0}}\}}\|\widehat{c}(V;\gamma)-c(V;\gamma)\|_{L_{2}}\right\} \\
 & \times\left\{ \sum_{\gamma\in\{\gamma_{R_{0}},\gamma_{S},\gamma_{S}+\gamma_{R_{0}}\}}\|\widehat{c}(V;\gamma)-c(V;\gamma)\|_{L_{2}}+\sum_{\gamma\in\{\gamma_{R_{0}},\gamma_{S},\gamma_{S}+\gamma_{R_{0}}\}}\|\widehat{b}(V;\gamma)-b(V;\gamma)\|_{L_{2}}\right\} .
\end{align*}
Let the probability
limits of the nuisance functions be $\mu_{a}^{*}(X)$, $\pi_{R_{s}}^{*}(X)$
and $\pi_{S}^{*}(X)$, respectively, we can show when $\gamma_S=\gamma_{R_1}=\gamma_{R_0}=0$ that the bias $P(S=1)(\widehat{\tau}^{[0]}-\tau)$ has the probability limit
\begin{align*}
P(S=1)(\widehat{\tau}^{[0]}-\tau) & \overset{p}{\rightarrow}P(S=1)\mathbb{P}\{\phi_{\text{eff}}^{^{*}[0]}(V)-\phi_{\text{eff}}^{[0]}(V)\}\\
 & =\mathbb{P}\left[\pi_{S}(X)\{1-\pi_{R_{1}}(X)\}\mu_{1}^{*}(X)+\pi_{S}(X)\pi_{R_{1}}(X)q_{R_{1}}^{*}(X)\{\mu_{1}(X)-\mu_{1}^{*}(X)\}\right]\\
 & -\mathbb{P}\left[\pi_{S}(X)\mu_{0}^{*}(X)+\{1-\pi_{S}(X)\}q_{S}^{*}(X)\frac{\pi_{R_{0}}(X)\{\mu_{0}(X)-\mu_{0}^{*}(X)\}}{\pi_{R_{0}}^{*}(X)}\right]\\
 & -\mathbb{P}\left[\pi_{S}(X)\{1-\pi_{R_{1}}(X)\}\mu_{1}(X)\right]+\mathbb{P}\left\{ \pi_{S}(X)\mu_{0}(X)\right\} \\
 & =\mathbb{P}\left(\pi_{S}(X)\{\mu_{1}(X)-\mu_{1}^{*}(X)\}\left[\pi_{R_{1}}(X)q_{R_{1}}^{*}(X)-\{1-\pi_{R_{1}}(X)\}\right]\right)\\
 & +\mathbb{P}\left(\{\mu_{0}(X)-\mu_{0}^{*}(X)\}\left[\pi_{S}(X)-\frac{\pi_{R_{0}}(X)\{1-\pi_{S}(X)\}q_{S}^{*}(X)}{\pi_{R_{0}}^{*}(X)}\right]\right)\\
 & =\mathbb{P}\left[\frac{\pi_{S}(X)\{\mu_{1}(X)-\mu_{1}^{*}(X)\}\{\pi_{R_{1}}(X)-\pi_{R_{1}}^{*}(X)\}}{\pi_{R_{1}}^{*}(X)}\right]\\
 & +\mathbb{P}\left[\frac{\{1-\pi_{S}(X)\}\{\mu_{0}(X)-\mu_{0}^{*}(X)\}\{q_{S}(X)\pi_{R_{0}}^{*}(X)-q_{S}^{*}(X)\pi_{R_{0}}(X)\}}{\pi_{R_{0}}^{*}(X)}\right].
\end{align*}
By the Cauchy-Schwarz inequality,
the remainder term for $\widehat{\tau}^{[0]}$ follows for some constants
$C_{1}$ and $C_{2}$, such that
\begin{align*}
\|{\rm Rem}(\widehat{P},P_{0})\|_{L_{2}} & =C_{1}\|\widehat{\pi}_{R_{1}}(X)-\pi_{R_{1}}(X)\|_{L_{2}}\times\|\widehat{\mu}_{1}(X)-\mu_{1}(X)\|_{L_{2}}\\
 & +C_{2}\left\{ \|\widehat{\pi}_{S}(X)-\pi_{S}(X)\|_{L_{2}}+\|\widehat{\pi}_{R_{0}}(X)-\pi_{R_{0}}(X)\|_{L_{2}}\right\} \times\|\widehat{\mu}_{0}(X)-\mu_{0}(X)\|_{L_{2}}.
\end{align*}
From another perspective, when $\gamma_{S}=\gamma_{R_{1}}=\gamma_{R_{0}}=0$,
we have $\|\widehat{c}(V;\gamma)-c(V;\gamma)\|_{L_{2}}=0$ for any
$\gamma$, 
\begin{align*}
\|\widehat{b}(X;\gamma_{R_{1}})-b(X;\gamma_{R_{1}})\|_{L_{2}} & =\|\widehat{\mu}_{1}(X)-\mu_{1}(X)\|_{L_{2}},\\
\sum_{\gamma\in\{\gamma_{R_{0}},\gamma_{S},\gamma_{S}+\gamma_{R_{0}}\}}\|\widehat{b}(V;\gamma)-b(V;\gamma)\|_{L_{2}} & =\|\widehat{\mu}_{0}(X)-\mu_{0}(X)\|_{L_{2}}.
\end{align*}
Substituting these bounds in the remainder term $\left[\int\{\phi_{\text{eff}}^{t}(V;\widehat{\mathbb{P}})-\phi_{\text{eff}}^{t}(V;P_{0})\}^{2}d\mathbb{P}\right]^{1/2}$,
we have 
\begin{align*}
\left[\int\{\phi_{\text{eff}}^{t}(V;\widehat{\mathbb{P}})-\phi_{\text{eff}}^{t}(V;P_{0})\}^{2}d\mathbb{P}\right]^{1/2} & \lesssim\|\widehat{q}_{R_{1}}(X)-q_{R_{1}}(X)\|_{L_{2}}\times\|\widehat{\mu}_{1}(X)-\mu_{1}(X)\|_{L_{2}}\\
 & +\|\widehat{\mu}_{0}(X)-\mu_{0}(X)\|_{L_{2}}\\
 & \times\left\{ \|\widehat{q}_{R_{0}}(X)-q_{R_{0}}(X)\|_{L_{2}}+\|\widehat{q}_{S}(X)-q_{S}(X)\|_{L_{2}}\right\} ,
\end{align*}
which leads to a similar conclusion for the primary analysis when
$\gamma_{S}=\gamma_{R_{1}}=\gamma_{R_{0}}=0$.

\subsection{Proof of Theorem \ref{thm:EIF_J2R}}

Under Model \ref{assump:tilting-MNAR-reduced},
the parameter $\tau$ can be identified by the tilting-J2R sensitivity
models:
\begin{align*}
\tau & =\E[\E[Y\{1,R(1)\}\mid X,S=1]\mid S=1]-\E[\E[Y\{0,R(0)\}\mid X,S=1]\mid S=1]\\
 & =\frac{1}{P(S=1)}\E(\pi_{S}(X)\E[Y\{1,R(1)\}\mid X,S=1]) \\
 & -\frac{1}{P(S=1)}\E(\pi_{S}(X)\E[Y\{0,R(0)\}\mid X,S=1]),\\
 & =\frac{1}{P(S=1)}\E\left\{ \pi_{S}(X)\pi_{R_{1}}(X)\mu_{1}(X)-\pi_{S}(X)\pi_{R_{1}}(X)\frac{d(X;\gamma_{R_{0}},\gamma_{S})}{e(X;\gamma_{R_{0}},\gamma_{S})}\right\} ,
\end{align*}
where
\begin{align*}
\E[Y\{1,R(1)\}\mid X,S=1]= & \E[R(1)Y(1,1)+\{1-R(1)\}Y(1,0)\mid X,S=1]\\
= & \E\{R(1)\mid X,S=1\}\E\{Y(1,1)\mid X,S=1,R(1)=1\}\\
 & +\E\{1-R(1)\mid X,S=1\}\E\{Y(1,0)\mid X,S=1,R(1)=0\}\\
= & \pi_{R_{1}}(X)\mu_{1}(X)+\{1-\pi_{R_{1}}(X)\}\E\big\{ Y(0,0)\mid X,R(0)=0,S=1\big\}\\
= & \pi_{R_{1}}(X)\mu_{1}(X)+\{1-\pi_{R_{1}}(X)\}\E\big\{ Y(0)\mid X,S=1\big\}\\
= & \pi_{R_{1}}(X)\mu_{1}(X)+\{1-\pi_{R_{1}}(X)\}\frac{d(X;\gamma_{R_{0}},\gamma_{S})}{e(X;\gamma_{R_{0}},\gamma_{S})},
\end{align*}
and 
\begin{align*}
\E[Y_{1}\{0,R(0)\}\mid X,S=1] & =\E[R(0)Y(0,1)+\{1-R(0)\}Y(0,0)\mid X,S=1]\\
 & =\E\big\{ Y(0)\mid X,S=1\big\}=\frac{d(X;\gamma_{R_{0}},\gamma_{S})}{e(X;\gamma_{R_{0}},\gamma_{S})}.
\end{align*}
To derive the EIF under the J2R assumption, we first derive the EIF
for the numerator, denoted by $\psi_{N_{\theta}}^{{\rm J2R}}(V)$
\begin{align*}
\psi_{N_{\theta}}^{{\rm J2R}}(V) & =\dot{\pi}_{S}(X)\pi_{R_{1}}(X)+\pi_{S}(X)\dot{\pi}_{R_{1}}(X)+\pi_{S}(X)\pi_{R_{1}}(X)\dot{\mu}_{1}(X)+\pi_{S}(X)\pi_{R_{1}}(X)\mu_{1}^{1}(X)\\
 & -\dot{\pi}_{S}(X)\pi_{R_{1}}(X)\frac{d(X;\gamma_{R_{0}},\gamma_{S})}{e(X;\gamma_{R_{0}},\gamma_{S})}-\pi_{S}(X)\dot{\pi}_{R_{1}}(X)\frac{d(X;\gamma_{R_{0}},\gamma_{S})}{e(X;\gamma_{R_{0}},\gamma_{S})}\\
 & -\pi_{S}(X)\pi_{R_{1}}(X)\frac{\partial}{\partial\theta}\left\{ \frac{d(X;\gamma_{R_{0}},\gamma_{S})}{e(X;\gamma_{R_{0}},\gamma_{S})}\right\} -\pi_{S}(X)\pi_{R_{1}}(X)\frac{d(X;\gamma_{R_{0}},\gamma_{S})}{e(X;\gamma_{R_{0}},\gamma_{S})}.
\end{align*}
Following similar proofs for Theorem \ref{thm:EIF-tilting}, we have
\begin{align*}
\psi_{N_{\theta}}^{{\rm J2R}}(V) & =SRY-\left\{ SR\frac{d(X;\gamma_{R_{0}},\gamma_{S})}{e(X;\gamma_{R_{0}},\gamma_{S})}+(1-S)q_{S}(X)\pi_{R_{1}}(X)h(V;\gamma_{R_{0}},\gamma_{S})\right\},
\end{align*}
and the corresponding EIF for $\tau$ under the tilting-J2R sensitivity model \ref{assump:tilting-MNAR-reduced} is 
\begin{align*}
\phi_{\text{eff}}^{t}(V) & =\frac{SRY}{P(S=1)}-\frac{1}{P(S=1)}\left\{ SR\frac{d(X;\gamma_{R_{0}},\gamma_{S})}{e(X;\gamma_{R_{0}},\gamma_{S})}+(1-S)q_{S}(X)\pi_{R_{1}}(X)h(V;\gamma_{R_{0}},\gamma_{S})\right\} \\
& -\frac{S}{P(S=1)}\tau,
\end{align*}
which completes the proof of Theorem \ref{thm:EIF_J2R}.

\subsection{Proof of Theorems \ref{Thm:eif0} and \ref{Thm:eif0-1}}
\begin{theorem}[EIF for the primary analysis]\label{Thm:eif0} Under
the assumptions in Theorem \ref{Thm:id_primary}, the nonparametric
EIF of $\tau$ is
\begin{align}
\phi_{\mathrm{eff}}^{[0]}(V;P_{0}) & =\frac{SRY}{P(S=1)}\nonumber \\
 & \ +\frac{S}{P(S=1)}\left[(1-R)\mu_{1}(X)+Rq_{R_{1}}(X)\{Y-\mu_{1}(X)\}\right]\label{eq:EIF0-part1}\\
 & \ -\frac{1}{P(S=1)}\left[S\mu_{0}(X)+(1-S)q_{S}(X)\frac{R\{Y-\mu_{0}(X)\}}{\pi_{R_{0}}(X)}\right]-\frac{S\tau}{P(S=1)}.\label{eq:EIF0-part2}
\end{align}
\end{theorem}
\begin{proof}
    Following the proof of Theorem \ref{thm:EIF-tilting}, let $\gamma_{S}=\gamma_{R_{1}}=\gamma_{R_{0}}=0$, we can show 
\begin{align*}
 & b(X;\gamma_{R_{1}})=\mu_{1}(X),\quad c(X;\gamma_{R_{1}})=1,\quad g(V;\gamma_{R_{1}})=Y-\mu_{1}(X),\\
 & d(X;\gamma_{R_{0}},\gamma_{S})=\mu_{0}(X),\quad e(X;\gamma_{R_{0}},\gamma_{S})=1,\\
 & h(V;\gamma_{R_{0}},\gamma_{S})=\frac{R\{Y-\mu_{0}(X)\}}{\pi_{R_{0}}(X)}.
\end{align*}
Plugging these term back to $\phi_{\text{eff}}^{t}(V)$ gives us the
EIF for the primary analysis: 
\begin{align*}
\phi_{\text{eff}}^{[0]}(V) & =\frac{S}{P(S=1)}RY\\
 & \ +\frac{S}{P(S=1)}\left[(1-R)\mu_{1}(X)+Rq_{R_{1}}(X)\{Y-\mu_{1}(X)\}\right]\\
 & \ -\frac{1}{P(S=1)}\left[S\mu_{0}(X)+(1-S)q_{S}(X)\frac{R\{Y-\mu_{0}(X)\}}{\pi_{R_{0}}(X)}\right]-\frac{S\tau}{P(S=1)},
\end{align*}
which completes the proof in Theorem \ref{Thm:eif0}.
\end{proof}

\begin{theorem}\label{Thm:eif0-1} For a function $f(X)$ with a
generic random variable $X$, denote its $L_{2}$-norm as $\|f(X)\|_{L_{2}}=\{\int f(x)^{2}dP(x)\}^{1/2}$.
Under the assumptions in Theorem \ref{Thm:eif0} and other regularity
conditions in Assumption \ref{assump:regularity}, we have 
\begin{align}
\widehat{\tau}^{[0]} & =\tau+\frac{1}{N}\sum_{i\in\mathcal{R}\cup\mathcal{E}}\phi_{\mathrm{eff}}^{[0]}(V_{i};P_{0})+\|{\rm Rem}(\widehat{P},P_{0})\|_{L_{2}}+o_{\pr}(N^{-1/2}),\label{eq:asym-EIF-tilting-decompose}
\end{align}
where $N=N_{\mathcal{R}}+N_{\mathcal{E}}$, $V_{\mathrm{eff}}^{[0]}=\E\{\phi_{\mathrm{eff}}^{[0]}(V;P_{0})\}^{2}$,
and
\begin{align*}
\|{\rm Rem}(\widehat{P},P_{0})\|_{L_{2}} & =C_{1}\|\widehat{\pi}_{R_{1}}(X)-\pi_{R_{1}}(X)\|_{L_{2}}\times\|\widehat{\mu}_{1}(X)-\mu_{1}(X)\|_{L_{2}}\\
 & +C_{2}\left\{ \|\widehat{\pi}_{S}(X)-\pi_{S}(X)\|_{L_{2}}+\|\widehat{\pi}_{R_{0}}(X)-\pi_{R_{0}}(X)\|_{L_{2}}\right\} \times\|\widehat{\mu}_{0}(X)-\mu_{0}(X)\|_{L_{2}},
\end{align*}
for some constants $C_{1}$ and $C_{2}$. 
\end{theorem}
\begin{proof}
  Since $\phi_{\mathrm{eff}}^{[0]}(V;P_{0})$ has mean zero, it motivates the construction of a more principled estimator by solving its empirical mean under the estimated distribution $\widehat{P}$ of the observed data:
\begin{align*}
\widehat{\tau}^{[0]} & = \frac{1}{N_{\mathcal{R}}}\sum_{i\in\mathcal{R}}\left[R_{i}\widehat{\mu}_{1}(X_{i})+R_{i}\{Y_{i}-\widehat{\mu}_{1}(X_{i})\}\right]\\
 & \quad + \frac{1}{N_{\mathcal{R}}}\sum_{i\in\mathcal{R}}(1-R_{i})\widehat{\mu}_{1}(X_{i})+\frac{1}{N_{\mathcal{R}}}\sum_{i\in\mathcal{R}}R_{i}\widehat{q}_{R_{1}}(X_{i})\{Y_{i}-\widehat{\mu}_{1}(X_{i})\}\\
 & \quad - \frac{1}{N_{\mathcal{R}}}\sum_{i\in\mathcal{R}}\widehat{\mu}_{0}(X_{i})-\frac{1}{N_{\mathcal{E}}}\sum_{i\in\mathcal{E}}\frac{R_{i}\widehat{q}_{S}(X_{i})}{\widehat{\pi}_{R_{0}}(X_{i})}\{Y_{i}-\widehat{\mu}_{0}(X_{i})\},
\end{align*}
where $\widehat{P}$ is characterized by the estimated nuisance models $\widehat{\mu}_{a}(X)$, $\widehat{\pi}_{R_{s}}(X)$, and $\widehat{\pi}_{S}(X)$. Assuming these converge to $\mu_{a}^{*}(X)$, $\pi_{R_{s}}^{*}(X)$, and $\pi_{S}^{*}(X)$, respectively, we can show that the bias $P(S=1)(\widehat{\tau}^{[0]}-\tau)$ has the probability limit $B_{1}+B_{2}$, where 
\begin{align*}
B_{1} & = \E\left[\frac{\pi_{S}(X)\{\mu_{1}(X)-\mu_{1}^{*}(X)\}\{\pi_{R_{1}}(X)-\pi_{R_{1}}^{*}(X)\}}{\pi_{R_{1}}^{*}(X)}\right],\\
B_{2} & = \E\left[\frac{\{1-\pi_{S}(X)\}\{\mu_{0}(X)-\mu_{0}^{*}(X)\}\{q_{S}(X)\pi_{R_{0}}^{*}(X)-q_{S}^{*}(X)\pi_{R_{0}}(X)\}}{\pi_{R_{0}}^{*}(X)}\right],
\end{align*}
which is the second remainder term $\|\mathrm{Rem}(\widehat{P},P_{0})\|_{L_{2}}$. From the proof of Theorem \ref{thm:EIF-tilting-1}, we have 
\begin{align*}
\|\mathrm{Rem}(\widehat{P},P_{0})\|_{L_{2}} & = C_{1}\|\widehat{\pi}_{R_{1}}(X)-\pi_{R_{1}}(X)\|_{L_{2}}\times\|\widehat{\mu}_{1}(X)-\mu_{1}(X)\|_{L_{2}}\\
 & \quad + C_{2}\left\{ \|\widehat{\pi}_{S}(X)-\pi_{S}(X)\|_{L_{2}}+\|\widehat{\pi}_{R_{0}}(X)-\pi_{R_{0}}(X)\|_{L_{2}}\right\} \times\|\widehat{\mu}_{0}(X)-\mu_{0}(X)\|_{L_{2}}
\end{align*}
when $\gamma_{S}=\gamma_{R_{1}}=\gamma_{R_{0}}=0$. By the central limit theorem, $\widehat{\tau}^{[0]}$ is rate-doubly robust, root-n consistent, and achieves the semi-parametric efficiency bound $V_{\mathrm{eff}}^{[0]}=\E\{\phi_{\mathrm{eff}}^{[0]}(V;P_{0})\}^{2}$ if the remainder term $\|\mathrm{Rem}(\widehat{P},P_{0})\|_{L_{2}}=o_{\pr}(N^{-1/2})$. This condition is met if the nuisance models converge at rates faster than $N^{-1/4}$, which is satisfied by some machine learning methods \citep{kennedy2016semiparametric,bradic2019sparsity}.
\end{proof}

\section{Extension: control-based imputation \label{sec:CBI}}

In this section, we assess the robustness of the primary analysis
with the control-based imputation (CBI) \citep{carpenter2013analysis}. This approach posits that the responses of participants who did not complete the treatment
in both groups are akin to the observed responses in the control group.
Inspired by the zero-dose model in \citet{little1996intent},
\citet{carpenter2013analysis} termed this approach as CBI and proposed several specific CBI methodologies,
including Jump-to-Reference (J2R), Copy Reference, Copy Increments in Reference, and Last
Mean Carried Forward. These methods characterize the relationship
between the missing outcomes in the treated group and the observed
outcomes in the control group. Diverging from MAR, CBI
is widely utilized in sensitivity analyses to assess the resilience
of study results against the untestable MAR hypothesis. It has been referenced in various studies (e.g., \citealp{carpenter2013analysis,yang2023smim,liu2024multiply,liu2024robust}),
and increasingly implemented to study clinical trials,
particularly in the domains of weight loss and chronic disease management, as noted by \citealp{tan2021review}.

\begin{figure}[htbp]
\centering
\includegraphics[width=.8\linewidth]{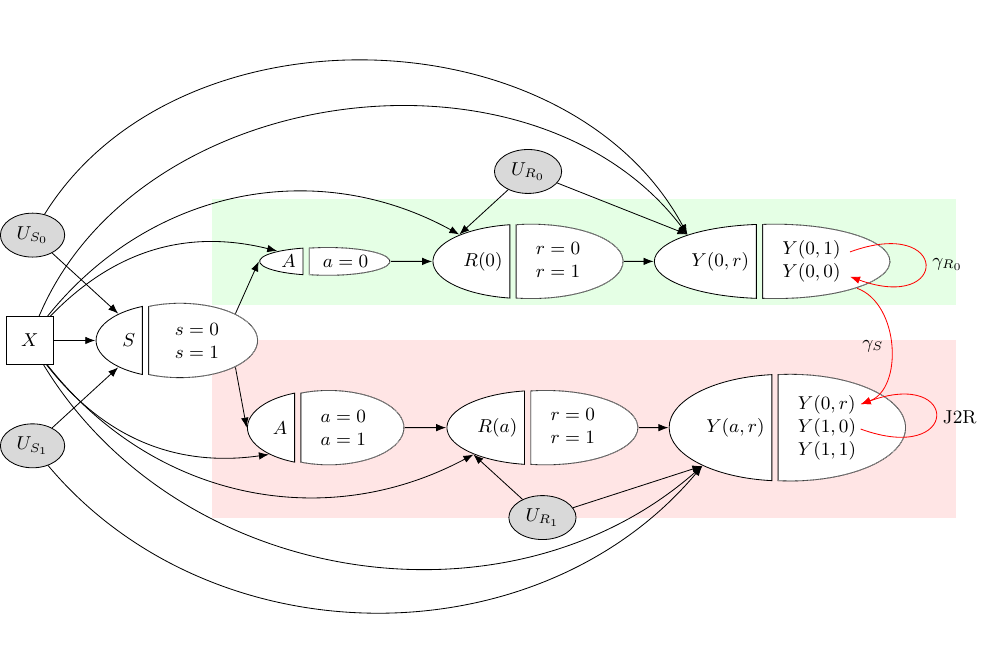}\caption{\label{fig:plot-DAG-J2R} Schematic plot of the J2R framework subject to unmeasured confounders
$U_{S}$, $U_{R_{0}}$ and $U_{R_{1}}$.}
\end{figure}

In particular, we primarily focus on the J2R framework as a special case of CBI methods under Model \ref{assump:tilting-MNAR-reduced}. In this framework, the treated participants in SAT are expected to have similar performance to those in the control group, given the same covariates, after the intercurrent event occurs. The J2R framework is widely applicable to trials for oncology and chronic diseases. Since it is common for patients in these trials to shift to standard care if they discontinue the test therapy for various reasons, the missing outcomes in the treatment group should parallel the outcome means of the control group with similar historical data; see Figure \ref{fig:plot-DAG-J2R} for an illustration when Assumptions 3 to 5 in Table \ref{tab:Key-assumptions} are violated due to unmeasured confounders $U_{S}$, $U_{R_{0}}$, and $U_{R_{1}}$ under the J2R framework.

% Partial ignorability of missingness and
\begin{model}[Modified tilting sensitivity models]\label{assump:tilting-MNAR-reduced}
Assume that the post-intercurrent event outcomes of the treatment group are $\E\left\{ Y(1,0)\mid X,R=0,S=1\right\} =\E\{Y(0)\mid X,S=1\}$. Assume that the tilting models for EC outcome
non-exchangeability and the effects of intercurrent events for external controls are
\begin{align*}
 & {\rm d}F\{Y(0)\mid X,S=0,R=0\}={\rm d}F\{Y(0)\mid X,S=0,R=1\}\frac{\exp\{\gamma_{R_{0}}Y(0)\}}{c(X;\gamma_{R_{0}})},\\
 & {\rm d}F\{Y(0)\mid X,S=1\}={\rm d}F\{Y(0)\mid X,S=0\}\frac{\exp\{\gamma_{S}Y(0)\}}{\E[\exp\{\gamma_{S}Y(0)\}\mid X,S=0]}.
\end{align*}
\end{model}
Model \ref{assump:tilting-MNAR-reduced} provides a methodological approach for estimating the conditional post-intercurrent
event outcome means in the SAT. Under this model, we have $\E\left\{ Y(1,0)\mid X,R=0,S=1\right\} =\mu_{0}(X)$, where the post-intercurrent event outcomes of the treatment group have the same conditional means as the control group. In contrast, Model \ref{assump:tilting-MNAR} relates the unobserved post-intercurrent event outcomes to those without intercurrent events through the exponential tilting model with the sensitivity parameter $\gamma_{R_1}$, rather than using control-based imputation. These two models represent different plausible scenarios for handling the outcome missingness due to the occurrence of intercurrent events. Next, we establish the non-parametric identification and develop the
EIF-motivated estimation of $\tau$ under the J2R framework in Theorem
\ref{thm:EIF_J2R}.

\begin{theorem}[Identification and EIF-tilting \& J2R models]\label{thm:EIF_J2R}

Under Assumptions 1 and 2 in Table \ref{tab:Key-assumptions}, and Model \ref{assump:tilting-MNAR-reduced} with fixed
$\gamma_{R_{0}}$ and $\gamma_{S}$, the following identification
formula and EIF holds for $\tau$:
\[
\tau=\frac{1}{P(S=1)}\E\left\{ \pi_{S}(X)\pi_{R_{1}}(X)\mu_{1}(X)-\pi_{S}(X)\pi_{R_{1}}(X)\frac{d(X;\gamma_{R_{0}},\gamma_{S})}{e(X;\gamma_{R_{0}},\gamma_{S})}\right\} ,
\]
and
\begin{align}
 &\phi_{\mathrm{eff}}^{\text{J2R}}(V;P_{0}, \gamma_{R_0}, \gamma_S) =\frac{SRY}{P(S=1)}\label{eq:EIF-observed-J2R}\\
 & -\frac{1}{P(S=1)}\left\{ \frac{SRd(X;\gamma_{R_{0}},\gamma_{S})}{e(X;\gamma_{R_{0}},\gamma_{S})}+(1-S)q_{S}(X)\pi_{R_{1}}(X)h(V;\gamma_{R_{0}},\gamma_{S})\right\} -\frac{S\tau}{P(S=1)}, \label{eq:EIF-gammaR0-gammaS-J2R} 
\end{align}
where $d(X;\gamma_{R_{0}},\gamma_{S})$, $e(X;\gamma_{R_{0}},\gamma_{S})$
and $h(V;\gamma_{R_{0}},\gamma_{S})$ are defined similarly in Theorem \ref{thm:id_tilting}. 

\end{theorem}

The first part (\ref{eq:EIF-observed-J2R}) of the EIF $\phi_{\mathrm{eff}}^{\text{J2R}}(V;P_{0}, \gamma_{R_0}, \gamma_S)$
is the same as (\ref{eq:EIF-observed}), which is contributed by the
trial participants with no intercurrent events. The second part (\ref{eq:EIF-gammaR0-gammaS-J2R})
is different from (\ref{eq:EIF-gammaR0-gammaS}) as it absorbs the
impact of intercurrent events within the treatment group under J2R.
By solving the empirical mean of $\phi_{\mathrm{eff}}^{\text{J2R}}(V;P_{0}, \gamma_{R_0}, \gamma_S)$
with estimated nuisance functions, we obtain the EIF-motivated estimator
for the sensitivity analysis under the J2R framework: 
\begin{align*}
\widehat{\tau}^{\text{J2R}} & =\frac{1}{N_{\mathcal{R}}}\sum_{i\in\mathcal{R}}R_{i}Y_{i}-\frac{1}{N_{\mathcal{R}}}\sum_{i\in\mathcal{R}}R_{i}\frac{\widehat{d}(V_{i};\gamma_{R_{0}},\gamma_{S})}{\widehat{e}(V_{i};\gamma_{R_{0}},\gamma_{S})}-\frac{1}{N_{\mathcal{R}}}\sum_{i\in\mathcal{\mathcal{E}}}\widehat{q}_{S}(X_{i})\widehat{\pi}_{R_{1}}(X)\widehat{h}(V_{i};\gamma_{R_{0}},\gamma_{S}).
\end{align*}

\end{document}